\documentstyle [psfig,11pt]{article}
\begin{document}
\begin{center}
{\Large The backscattering of polarised light from turbid media -}\\
{\Large An analysis of the azimuthal intensity variations
and its implications \\for the position of the source of diffusing radiation}\\
\vspace*{0.4cm}
Venkatesh Gopal$^{1}$,Hema Ramachandran${^2}$ and A.K.Sood.$^{1,3}$ \\
\vspace*{0.2cm}
${^1}$Department of Physics, Indian Institute of Science, Bangalore
560 012, INDIA\\
${^2}$Raman Research Institute, Sadashivanagar, Bangalore 560 080,
INDIA \\
${^3}$Jawaharlal Nehru Centre for Advanced Scientific Research, \\
Jakkur Campus, Bangalore 560 064, INDIA
\end{center}

\begin{abstract}
We study the azimuthal variation in the backscattered intensity that
is seen when polarised light is scattered by a turbid medium. We
present experimental observations of these intensity variations in
colloidal suspensions over a wide range of optical densities. For a
medium composed of spherical scatterers, we have developed using Mie
scattering theory and Monte Carlo simulations of photon transport, a
model which calculates the constant intensity contours of the
backscattered intensity. Comparisons of calculated and experimentally
obtained contours show very good agreement. To our knowledge, this is
the first model to provide a quantitative comparison with experimental
data. Close to the exact backscattering direction, where we have made
our intensity measuements, we show that the patterns are formed by
what we have called `reflected snake photons'. These are photons that
have been backscattered once and have maintained their direction of
propagation thereafter until they exit the medium. We also find that
the reflected snake photons originate within a depth of about $4l^{*}$
from the point of entry of the incident beam, where $l^{*}$ is the
photon transport mean free path. Further, in a novel
approach, we have used these patterns as a probe of the assumptions
underlying the diffusion approximation and present new results on the
position of the apparent source of diffusing radiation within the
medium. Possible applications are also discussed.

\end{abstract}

\section{Introduction}

When polarised light is incident on a turbid medium such as a
colloidal suspension, an azimuthal variation in the backscattered
intensity is observed. For a medium composed of spherical scattering
particles, patterns with a two-fold or four-fold symmetry are seen,
depending on the relative orientations of the polariser and analyser
\cite{Bigio1,Bigio2,Johnson,Dogariu}. Such patterns are seen in
situations as diverse as scattering from clouds \cite{Carswell},
biological cell suspensions \cite{Bigio1, Bigio2, Johnson} and the
macular region of the retina \cite{Hochheimer}. It is therefore of
interest to develop a model whereby, given the pattern, one can deduce
the properties of the scattering medium that gives rise to them, thus
permitting a quantitative characterisation of the scattering medium
from a non-invasive measurement. This would also be particularly
useful for biomedical applications.

Marston \cite {Marston1} and Carswell and Pal \cite {Carswell} had
surmised correctly that these patterns must have their origin in the
single scattering characteristics of the scattering particles. They
showed using Mie theory, the existence of an azimuthal variation in
the backscattered intensity from a single particle, and were thus able
to qualitatively explain the origin of these patterns. Rakovi\'{c}
and Kattawar \cite{Rakovic}, assuming that the scattering of light was
incoherent, showed that for an unpolarised incident beam, these
patterns could also be obtained as a consequence of the double
scattering of light. In this approximation they were able to
qualitatively predict, using a model based on Mie theory with no
fitting parameters, the shape of the patterns but not the finer
details.

In an approach that is completely different from previous analyses of
this effect, we find that the problem admits of a much simpler
solution, for scattering close to the exact backscattering direction,
in terms of what we have called `reflected snake photons'. These are
snake photons that travel some distance into the medium, are
backscattered by a large angle, and once again travel almost
undeviated in the direction that they have been scattered until they
exit the sample. We present a model based on Mie theory, that is
capable of quantitatively predicting the shape of the constant
intensity contours, and gives excellent agreement with our
experimental results. Additionally, which to our knowledge has not
been recognised so far, we find that these patterns can be used to
test the assumptions underlying the popular diffusion approximation
that is commonly used in interpreting experiments involving the
multiple scattering of light \cite{Morse and Feshbach}. In this
regard, we obtain new results concerning the shape and position of the
apparent source of diffusing radiation within a random scattering
medium.

We digress briefly to outline the theory of multiple scattering theory
and define the relevant quantities and notation. We also confine our
attention in the rest of this paper to spherical dielectric
scatterers. On entering a scattering medium, photons travel
exponentially distributed ballistic pathlengths between scattering
events. The scattering mean free path $l_{s}$ is the mean distance
between scattering events and is determined by the scattering cross
section $\sigma$ and the number density $\phi$ of the scatterers as
$l_{s} = 1/\sigma \phi$. The scattering cross sections and the
probability to scatter at a given angle are calculated by the well
known Mie theory \cite{Bohren and Huffman}. The angular distribution
of scattered light depends on the size parameter $x = 2\pi
a/\lambda$, where $a$ is the particle radius and $\lambda$ the
wavelength of the incident light. For $a \ll \lambda$, the scatterers
may be approximated by dipoles and the scattering is well described by
Rayleigh scattering. When $a$ is comparable to or greater than
$\lambda$, interference effects arise and the scattering is peaked in
the forward direction \cite{Feynman} and the Mie theory must now be
employed. The result of this anisotropy in scattering is that the
photon is often not randomised after a single scattering event and
there is a `persistence length' over which the photon travels, on
average, in approximately the same direction before being randomised.
If the photon undergoes a very large number of scattering events, then
one may assume that the photon performs a random walk and that the
photon flux is transported diffusively within the medium. The
persistence length or the length scale over which the photon is
randomised is called the transport mean free path $l^{*}$
\cite{lstar}. The `optical thickness' or optical density $\tau$ of a
slab of thickness $L$ is defined as $\tau = L/l^{*}$. When $L \sim
l^{*}$ the scattering is largely ballistic, and, when $L \gg l^{*}$,
the diffusion approximation is valid \cite{Pine,lstar}. The
scattering anisotropy $g$, a measure of the persistence length, is
defined in terms of the mean free paths $l^{*}$ and $l_{s}$ as $g = 1
- (l_{s}/l^{*})$. Photons travelling through a random medium may be
classified into three types:  a ballistic component that has not
undergone any scattering, a diffuse component that is completely
randomised directionally and may be modelled by a diffusion equation,
and a quasi-ballistic or `snake' component, that has undergone more
than one scattering event but is still travelling in approximately the
same direction as it did when it entered the medium.

We describe our experimental setup and observations in sections 2 and
3. In section 4 we provide a first description of our model. This
model, which we term the `slice model', requires information about two
length scales. We then show, how conventional diffusion theory is
unable to provide these two parameters. In the absence of an
analytical framework for deriving these parameters from first principles,
we employ Monte Carlo random walk simulations to obtain these lengths.
Section 5 describes our simulation technique and the results obtained.
In section 6 we add further detail to the slice model by including the
results of the simulations and then compare our calculations with our
experimental data. Section 7 describes some shortcomings of our model
and how these may be remedied. The slice model, in conjunction with
random walk simulations of photon transport yields new insight into
the long standing problem of the position of the effective source of
diffusing photons. We present our results in this regard in section
8. Finally, we conclude in section 9, in which we also describe
briefly some possible applications. Some of the results described
here have been previously reported \cite{Venkatesh2}, but in order
that the paper be self-contained, we have reproduced them here as
well.

\section{Experimental details}

\begin{figure}[!htbp]
\centerline{\psfig{figure=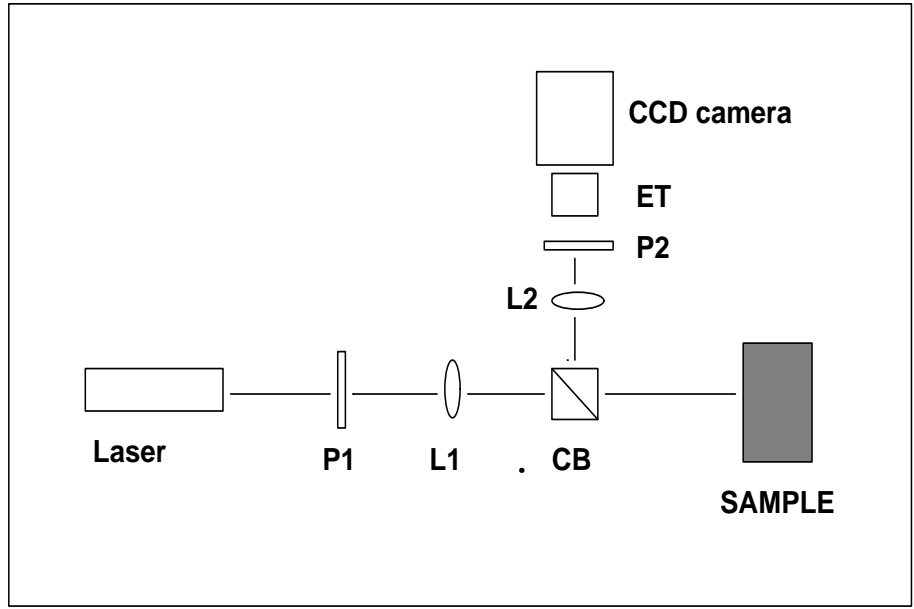,width=12cm,height=8cm}}
\caption{Schematic diagram of the experimental apparatus :  {\bf P1} :
Polariser, {\bf L1} :  Lens (f = 150mm), {\bf CB} :  Non-polarising
cube beam-splitter, {\bf L2} :  Lens (f = 50mm), {\bf P2} :  Analyser,
{\bf ET} :  Extender tube and camera lens.}
\end{figure}
A schematic diagram of the experimental setup used is shown in Fig.
1. Light from a randomly polarised He-Ne (Melles-Griot, $\lambda =
612nm$) laser was focussed using a 150 mm focal length lens {\bf L1},
onto a cuvette containing the scattering medium, an aqueous suspension
of 0.23 $\mu m$ diameter polystyrene spheres (Seradyn, U.S.A) of
varying volume fractions. The incident light was linearly polarised
using the polariser {\bf P1}. The backscattered light was imaged
using a non-polarising cube beam splitter and a 50mm focal length lens
{\bf L2}. In order to obtain a magnified image, extender tubes were
fitted in between the lens and the body of the CCD camera. The lens
{\bf L2} and the lens of the CCD camera together formed the imaging
optics. The polarisation to be viewed was selected by the analyser
{\bf P2} placed just before the camera. The intensified CCD camera
had a 512 $\times$ 512 pixel resolution and an 8 bit (0-256 grayscale)
intensity resolution capability. The sample cell was made of quartz
glass to prevent coagulation of the colloidal particles due to ionic
leaching from the walls. For each suspension, observations were made
with three relative orientations of the polariser and analyser. In
the first, only the polariser {\bf P1} was present (PO geometry) with
no analyser. In the second, the polariser {\bf P1} and analyser {\bf
P2} were aligned parallel to each other (VV geometry) while in the
third measurement they were crossed (VH geometry). For each
orientation, three images were recorded and averaged over.
Experiments were performed using suspensions of 5 different optical
densities. Details of these suspensions are given in Table 1. At the
wavelength of light used, the particles have a scattering anisotropy
$g = 0.452$.

\begin{table}
\begin{center}
\begin{tabular}{|c|c|c|c|c|c|} \hline
Optical density&0.51&1.84&3.56&8.76&17.22 \\
$\tau = L/l^{*}$&&&&& \\ \hline
Volume fraction &0.00011& 0.00041 & 0.0008 & 0.00196 & 0.00386 \\
$\phi$&&&&& \\ \hline
$l_{s}$ (mm)&10.63&2.97&1.54&0.63&0.32 \\ \hline
$l^{*}$ (mm)&19.4&5.41&2.8&1.14&0.58 \\ \hline
\end{tabular}
\end{center}
\caption{Optical density, volume fraction and mean free paths
of the scattering suspensions used in the experiment.}
\end{table}

\section{Experimental Results}

Figure 2 shows contour plots of the backscattered intensity patterns
obtained at different optical densities. The images in the first
column correspond to the PO geometry, the second column to VV and the
third column to VH, which are as described above. At the lowest
optical density of 0.51, images for PO and VV could not be recorded
due to excessive stray light reflections. The image of the VH pattern
at $\tau = 0.51$ is shown in Fig. 4b instead. In all the figures,
one can see that with increasing optical density, the images approach
a smooth circularly symmetric pattern, which is due to the diffuse
intensity. However, it is interesting to note that even with a sample
which is very turbid with $\tau = 17.22$, the azimuthal variations are
still seen, albeit faintly. Thus, we may guess that the patterns are
due to scattering very close to the back face and that the diffuse
intensity makes no contribution to these patterns other than to round
out the sharp intensity contours, a fact that has also been borne out
by our simulations and which has been used to estimate the position of
the source of diffusing photons.

\begin{figure}
\centerline{\psfig{figure=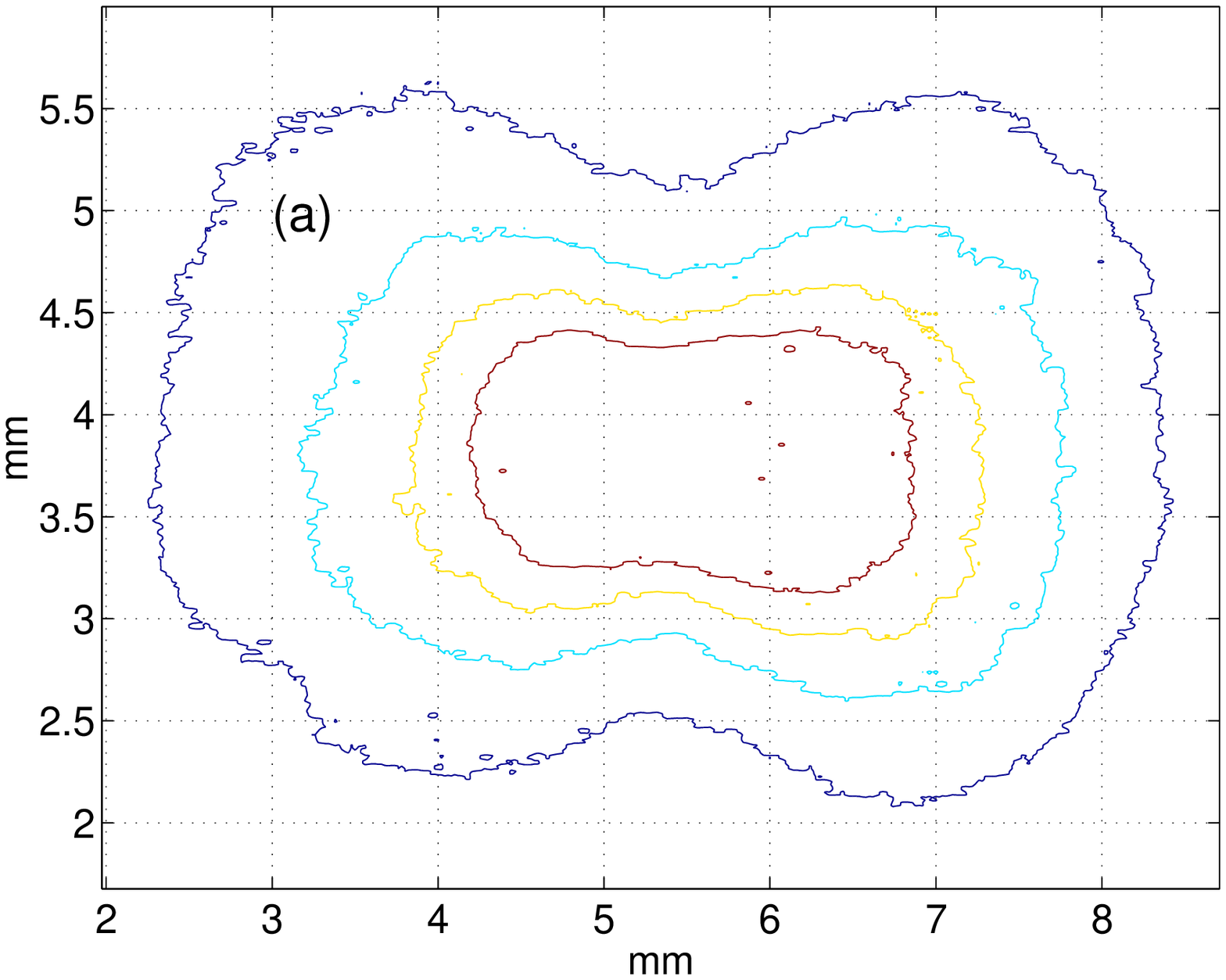,width=4.5cm,height=4.5cm}
\psfig{figure=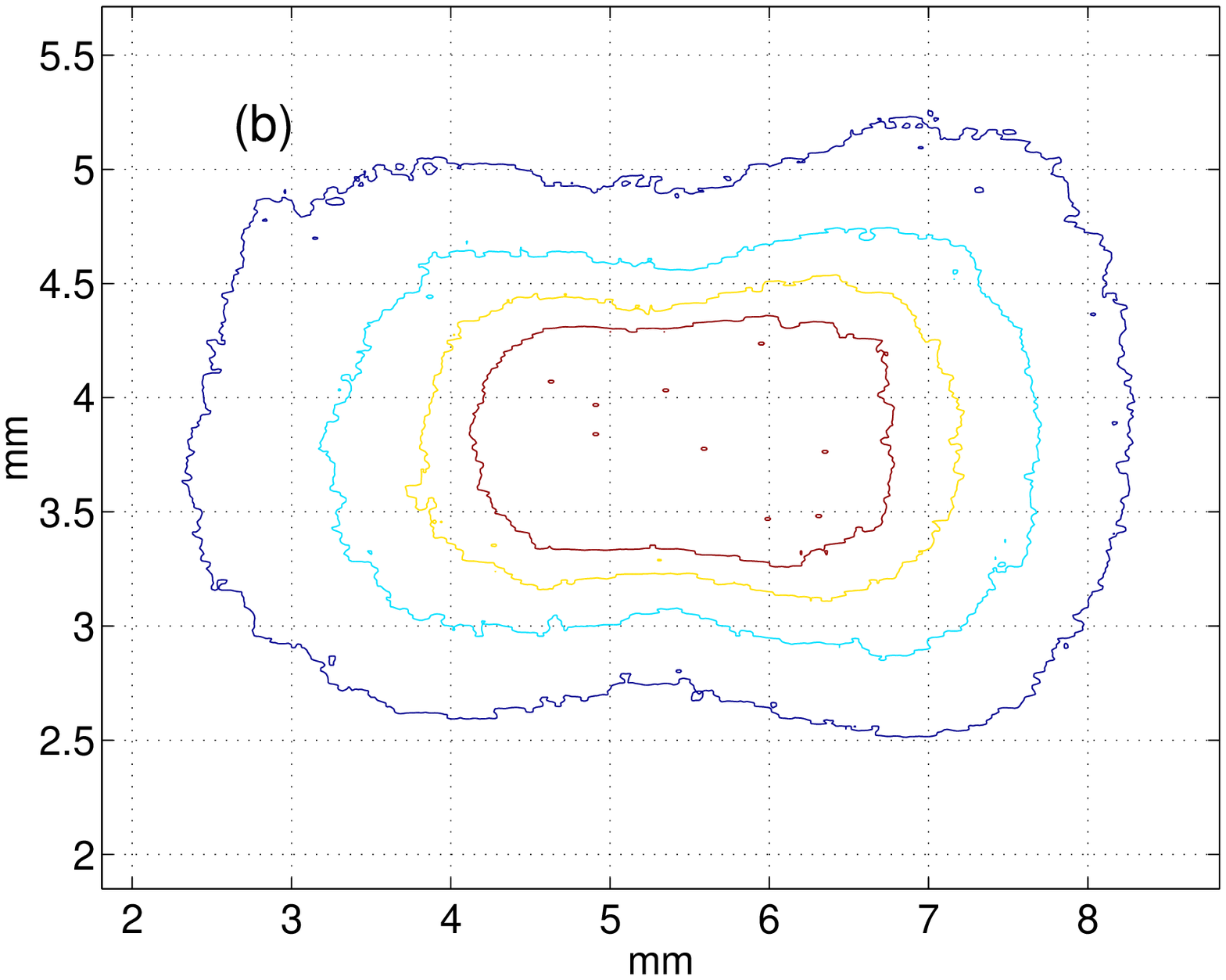,width=4.5cm,height=4.5cm}
\psfig{figure=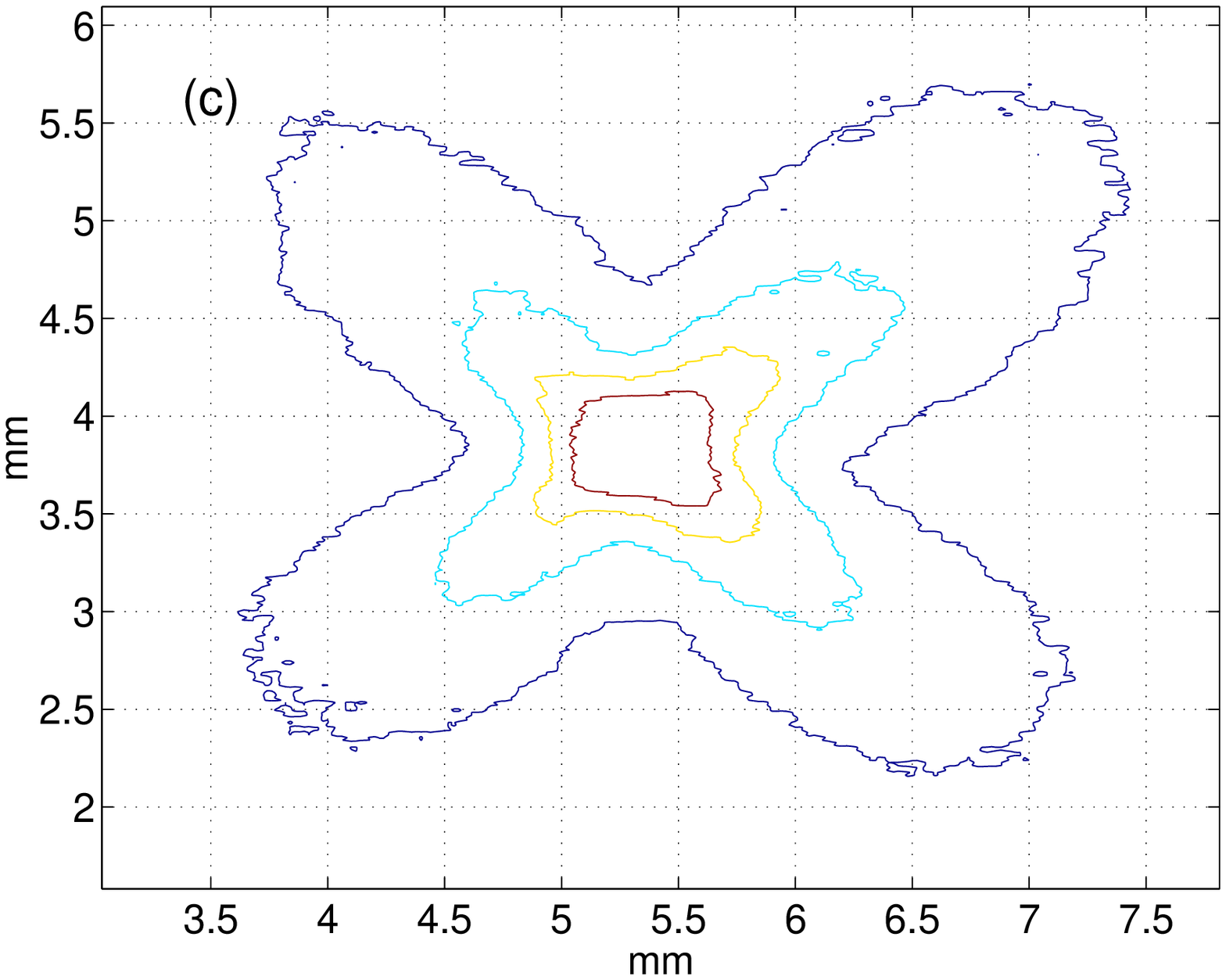,width=4.5cm,height=4.5cm}}
\centerline{\psfig{figure=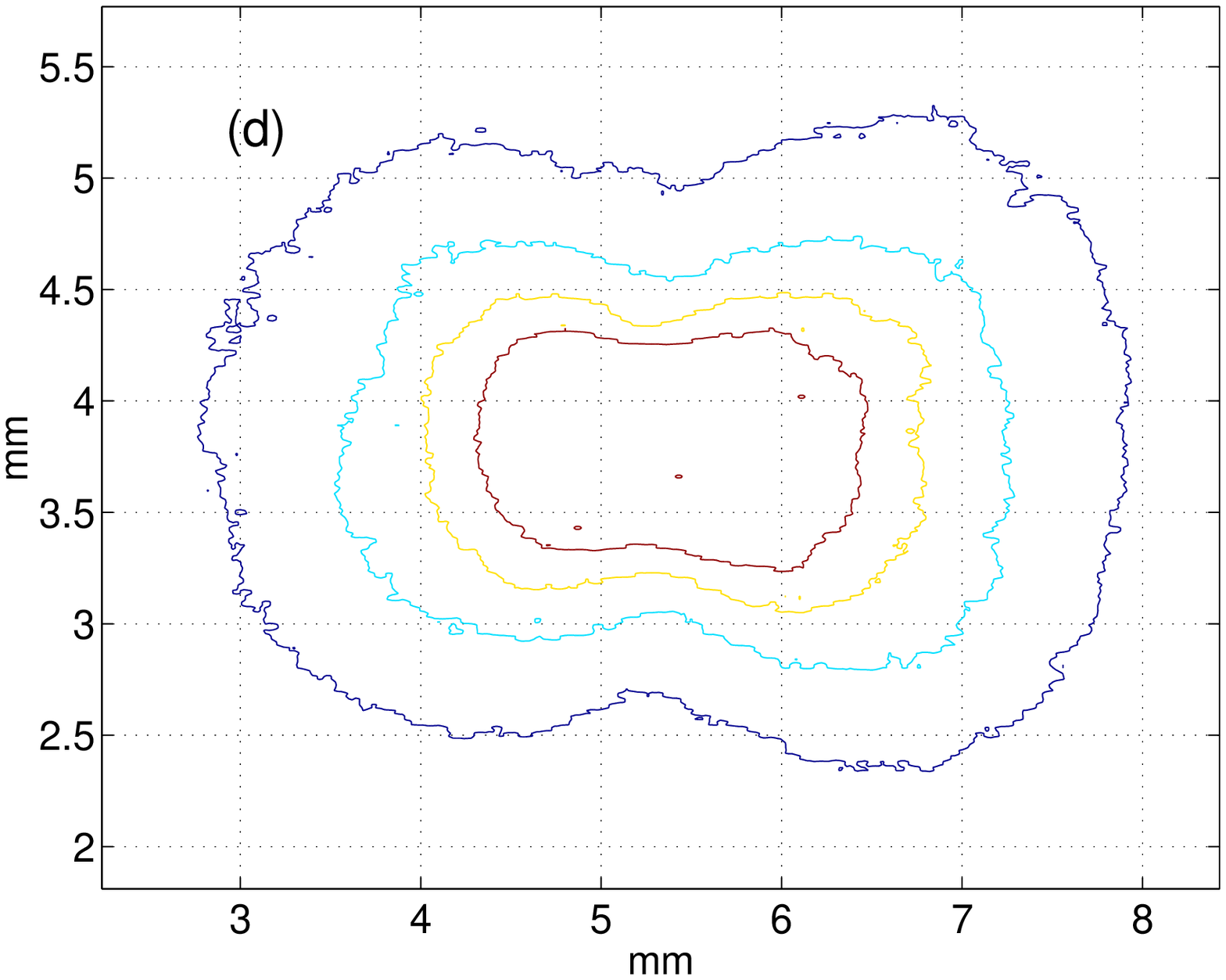,width=4.5cm,height=4.5cm}
\psfig{figure=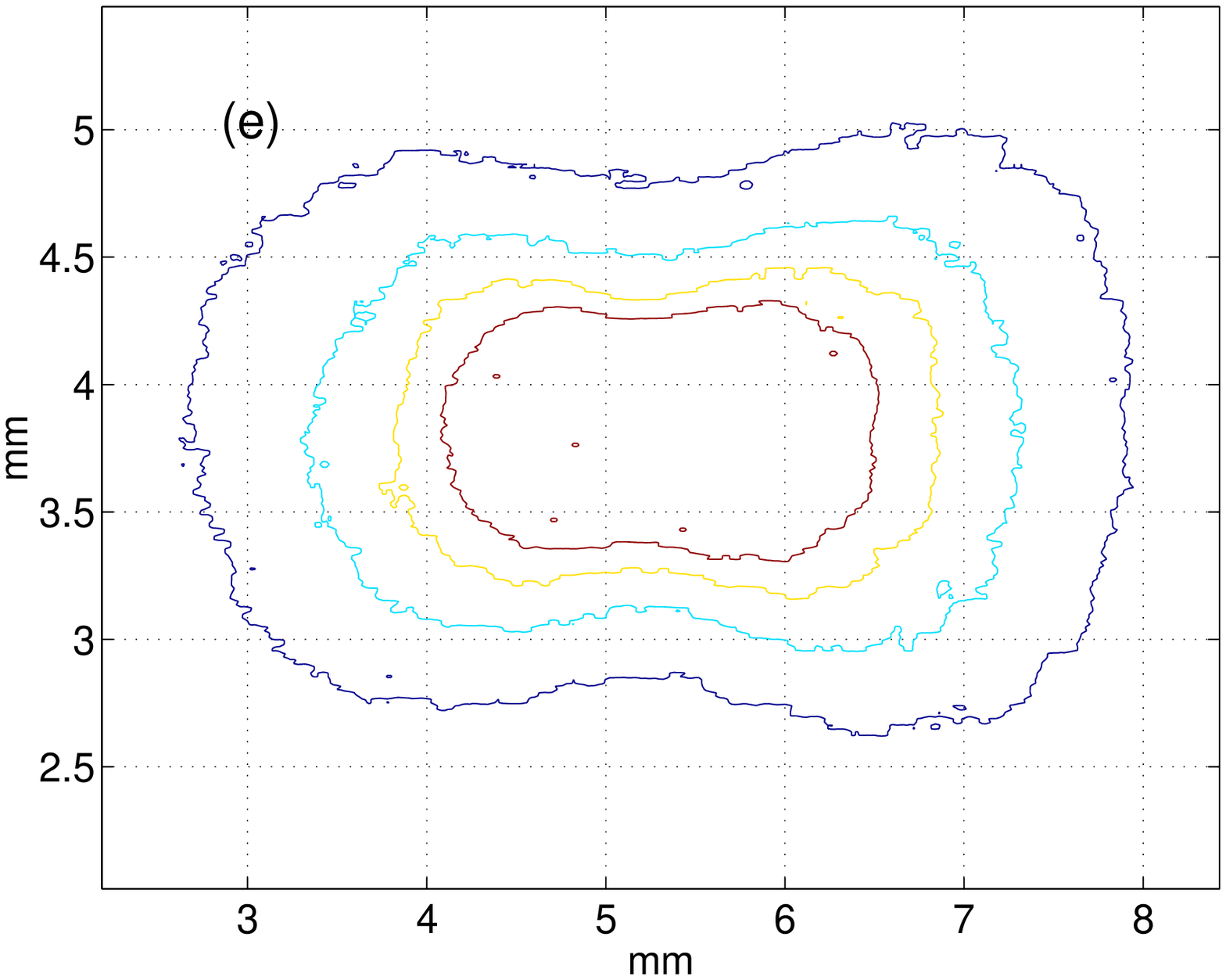,width=4.5cm,height=4.5cm}
\psfig{figure=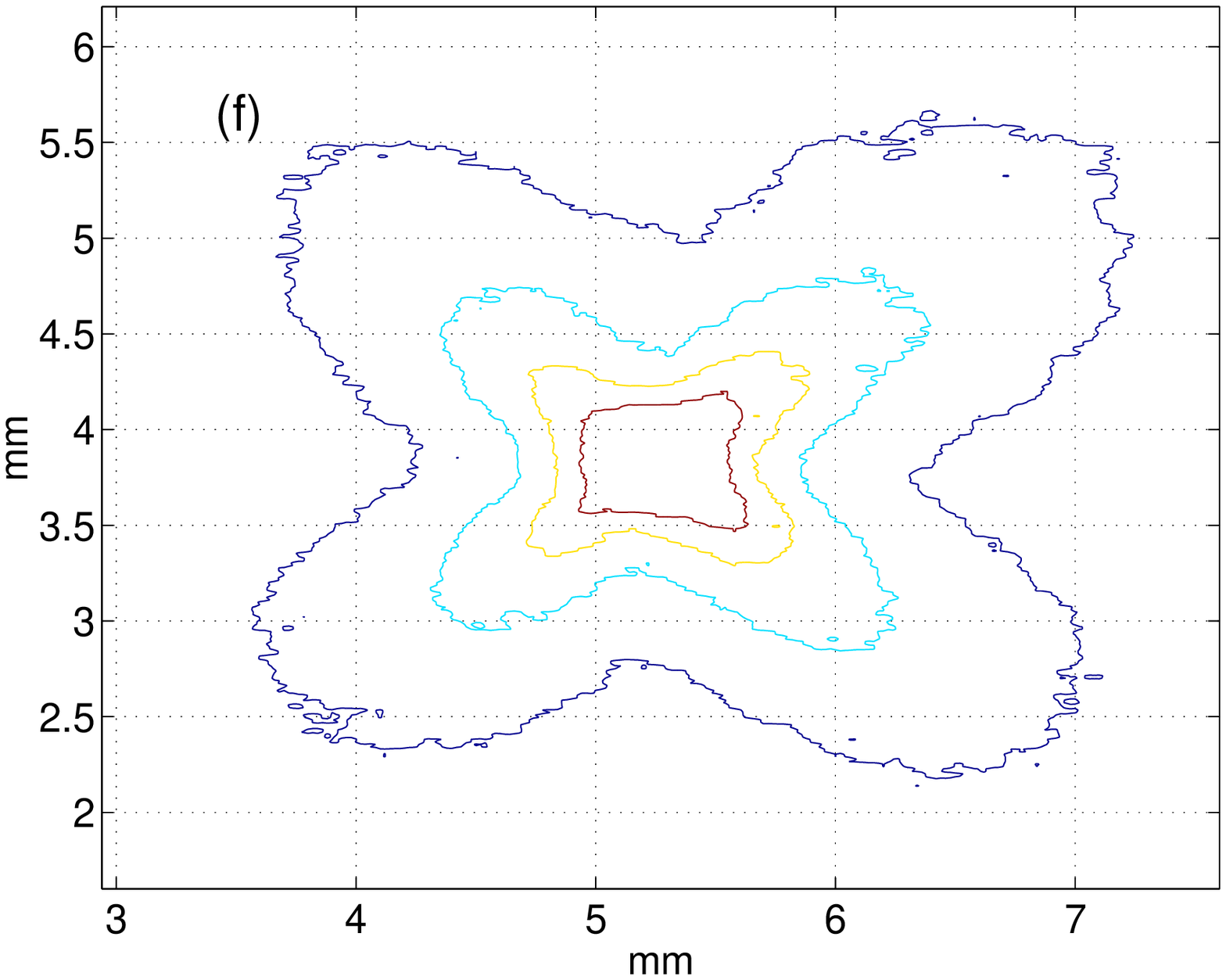,width=4.5cm,height=4.5cm}}
\centerline{\psfig{figure=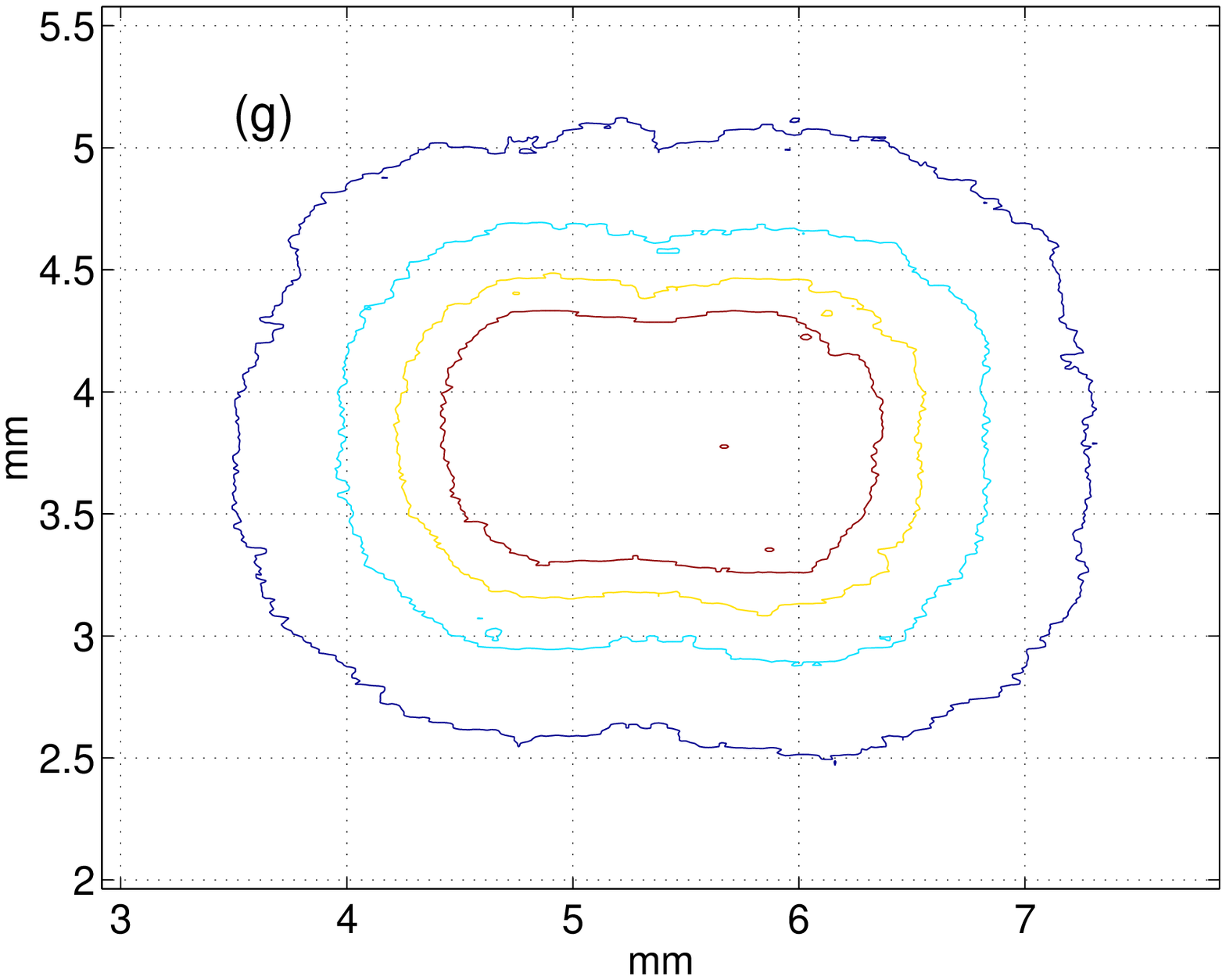,width=4.5cm,height=4.5cm}
\psfig{figure=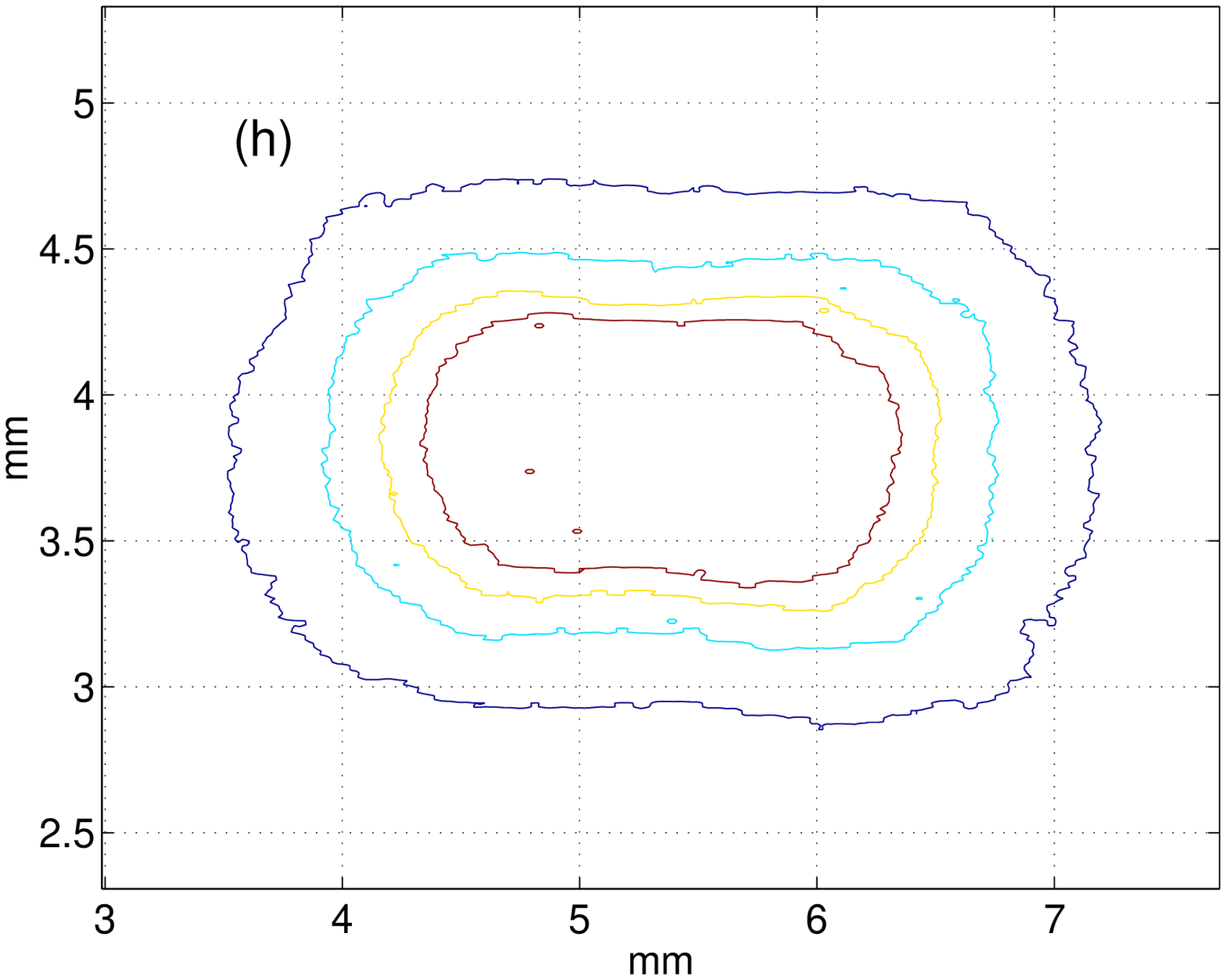,width=4.5cm,height=4.5cm}
\psfig{figure=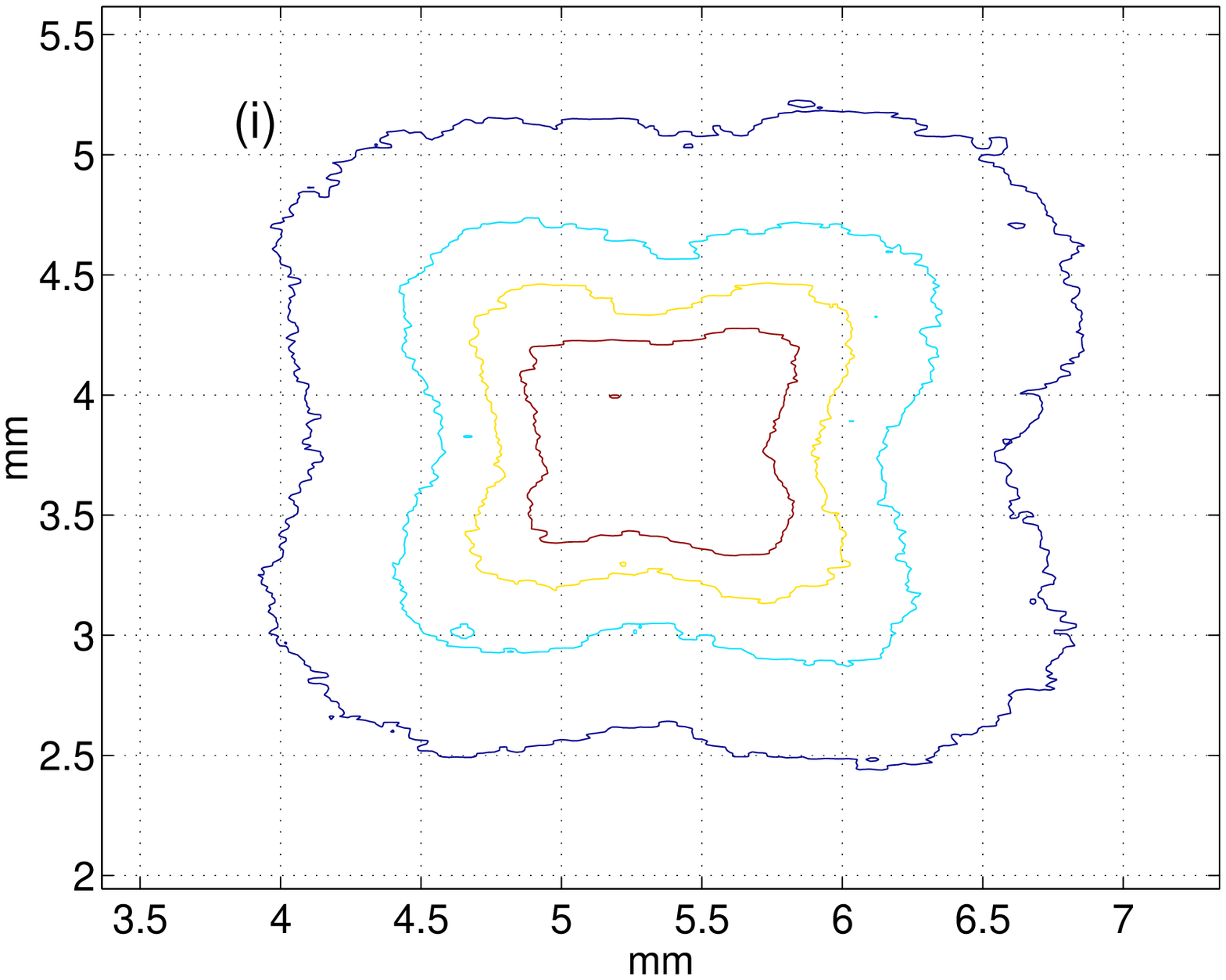,width=4.5cm,height=4.5cm}}
\centerline{\psfig{figure=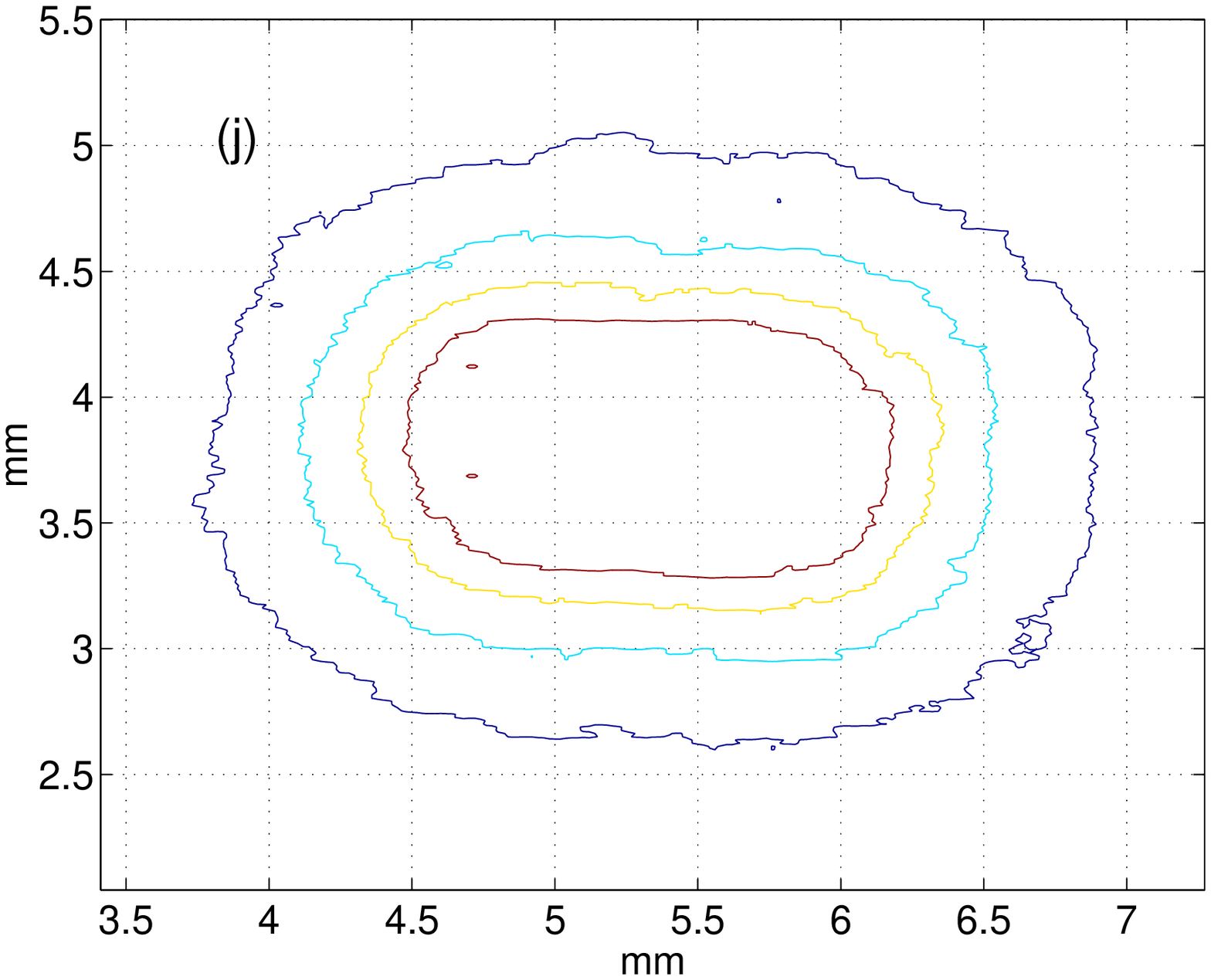,width=4.5cm,height=4.5cm}
\psfig{figure=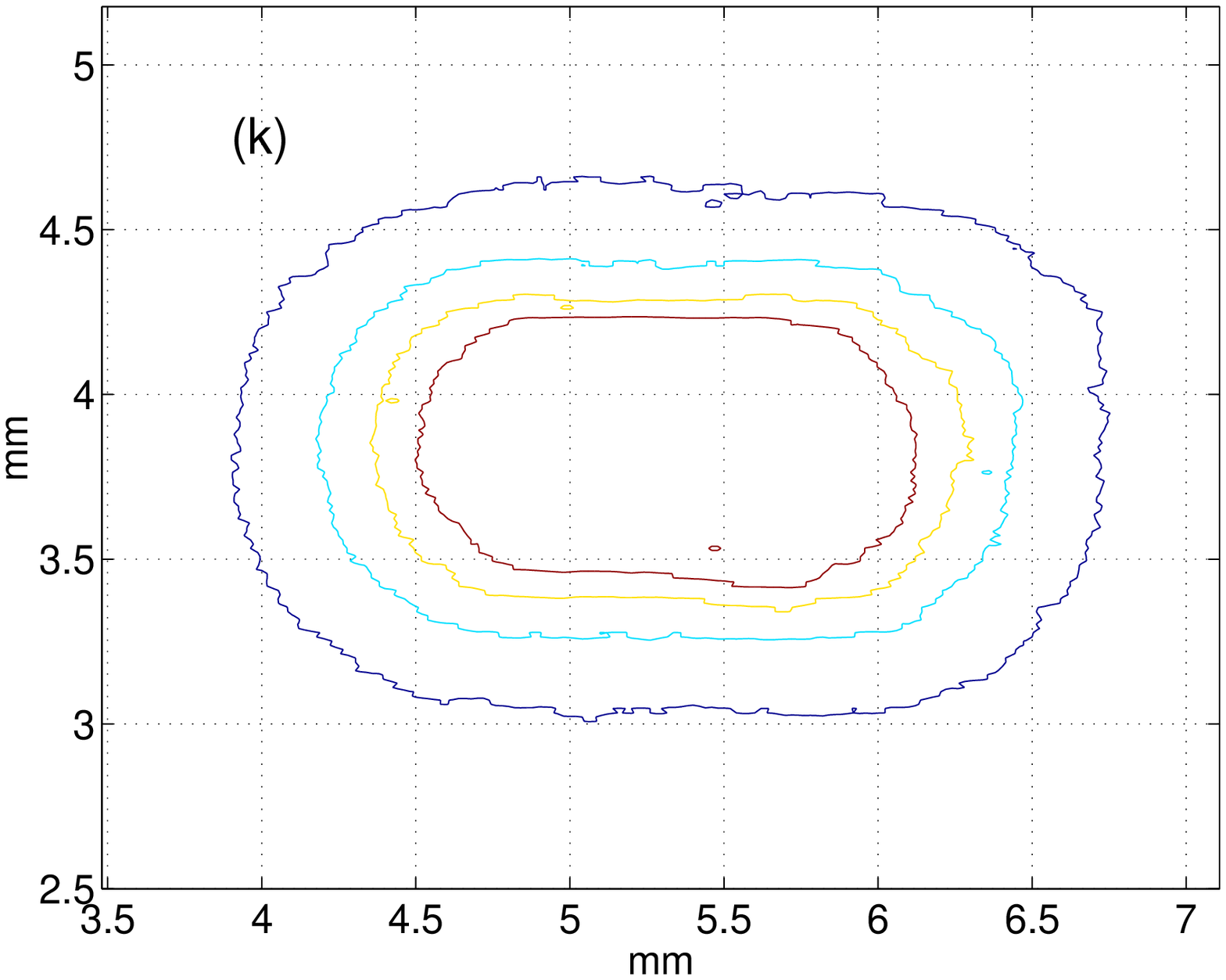,width=4.5cm,height=4.5cm}
\psfig{figure=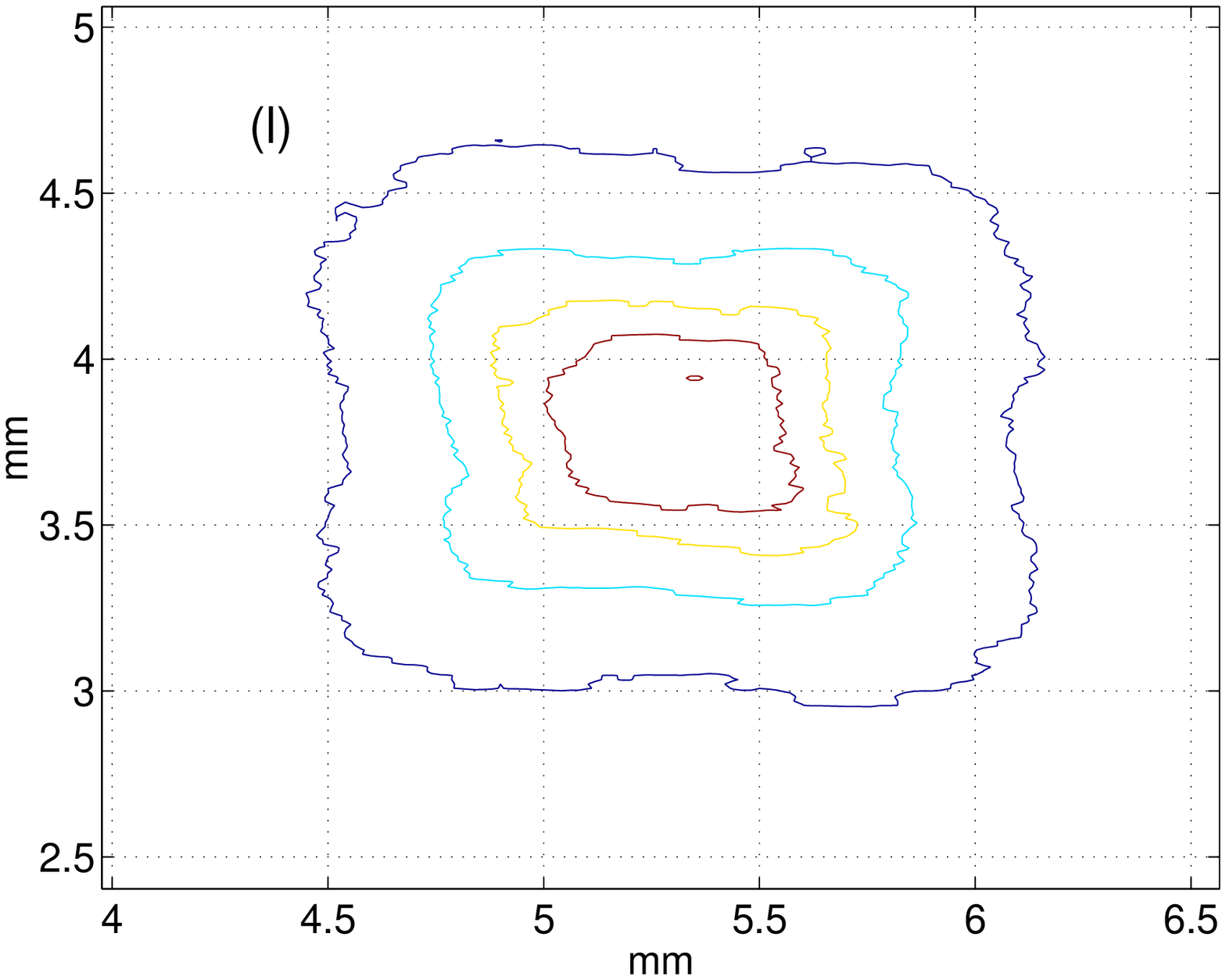,width=4.5cm,height=4.5cm}}
\caption{Contour plots of the backscattered intensity obtained at varying
optical densities. The images in the first column correspond to the
PO geometry, the second column to VV and the third to the VH geometry.
The figures are measured at the following optical densities :
a,b and c, $\tau = 1.86$; d, e and f, $\tau = 3.56$, g, h and i, $\tau = 8.76$;
 j, k and l, $\tau = 17.22$. Four evenly spaced inetnsity 
values have been chosen for all the contours.} 
\end{figure}
\section{The `slice' model}

Consider, as shown in Fig. 3, a thin beam of polarised light entering
along the $z$ axis, 
a random medium consisting of spherical scatterers.
Assuming that photons undergo scattering events on the scale of the
scattering mean free path, we divide the medium into $N$ `slices' by
means of planes parallel to the $x-y$ plane, two of which are shown in Fig. 3.
These planes are separated by a length $s$ which is of the order of
$l_{s}$. In the following discussion, when we speak of the $n$th
slice, we shall mean the region bounded by the $n$ and $(n+1)$th
planes. The input face, or back face of the cuvette where the light
is incident, is the plane with $n = 1$. Each slice has associated
with it a source, marked $S_{n}$(n = 1,2..N) in Fig. 3, which lies at
the centre of the $(n+1)$th plane. All the scattering that takes
place between the planes $n$ and $n+1$, is lumped together and assumed
to scatter from the point $S_{n}$, the `source' for the $n$th slice
which lies on the $(n+1)$th plane. Thus the transport of light within
the medium is approximated by a discrete model, which assumes that the
medium consists of a series of point sources of varying intensities
lying along the $z$ axis. Edge effects are avoided by assuming that
the transverse extent of the scattering medium is much greater than
its thickness. The slice model is essentially a discrete bookkeeping
scheme for the calculation of these source intensities.
\begin{figure}[!h]
\centerline{\psfig{figure=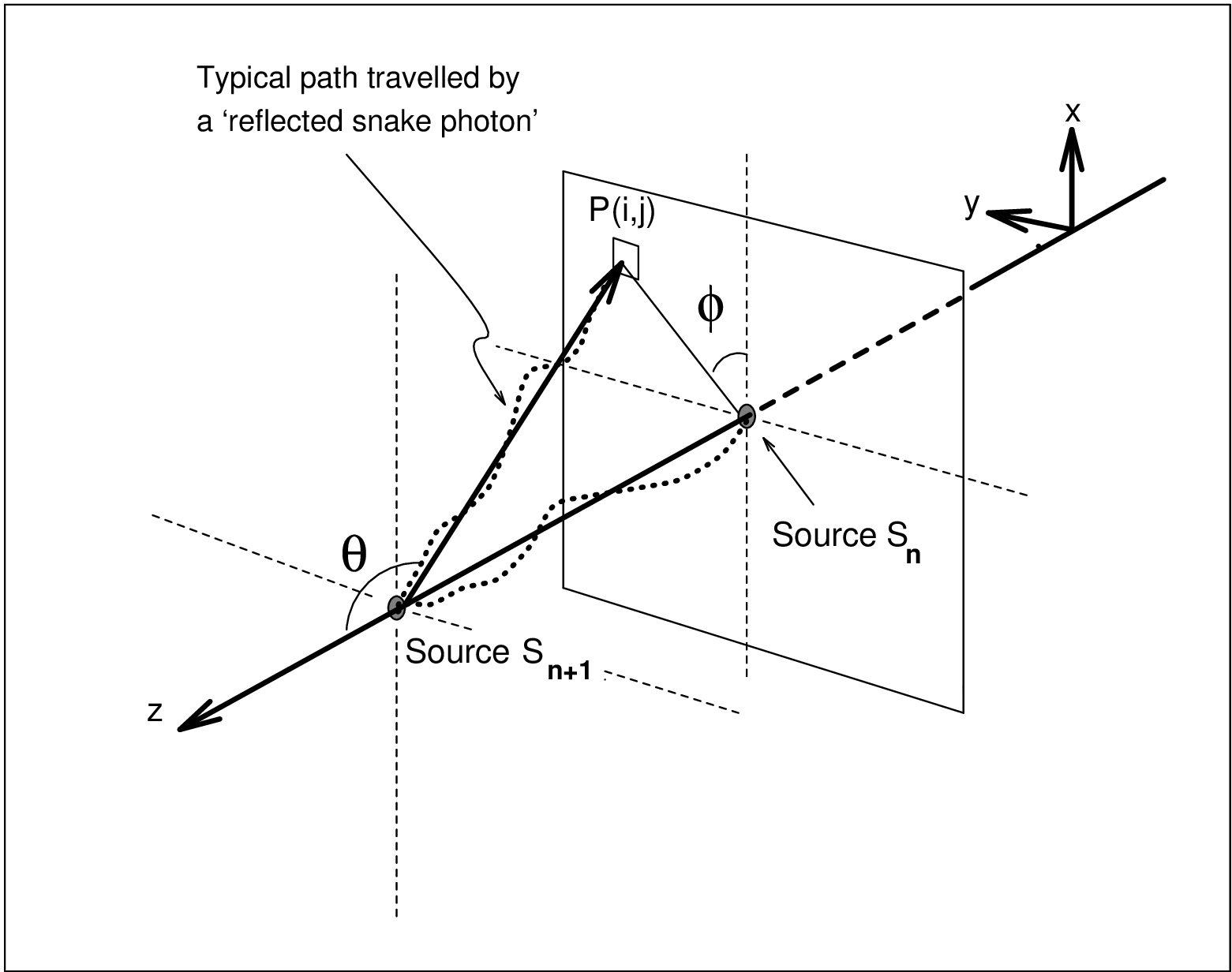,width=12cm,height=9cm}}
\caption{The geometry of the `slice' model. A beam of polarised light is
incident on the scattering medium along the $+z$ direction. All the
intensity scattered between the $n$th and the $(n+1)$th planes is
assumed to scatter from a source at the centre of the $(n+1)$th plane,
represented here by the source marked $S_{n}$. The $n$th slice is
bounded by the $n$th and $(n+1)$th planes. The incoming light is
polarised along the $x$ axis. The scattering patterns are obtained by
mapping the Mie scattering cross-section onto a 100 $\times$ 100 grid
of pixels superimposed on the back face of the scattering cell.
Assuming the plane containing the source $S_{n}$ to be the input face,
a pixel $P(i,j)$, whose centre subtends an angle $(\theta, \phi)$ at
the source $S_{n+1}$, is shown. The dotted line shows a typical path
of a `reflected snake photon' which is responsible for the azimuthal
intensity patterns.}
\end{figure}

The intensity of the source $S_{n}$, is the fraction of the incident
intensity that is scattered between the planes $n$ and $(n+1)$. This
fraction consists of two parts. The first is the ballistic part of
the incident flux consisting of photons which have travelled
unscattered upto the $n$th plane, and then scattered before reaching
the $(n+1)$th plane. The second is due to the `snake component',
photons that have previously been scattered but have remained largely
undeviated and have continued to travel along their initial direction
of propagation, and have been scattered once again between the $n$ and
$(n+1)$th planes. The ballistic intensity that is scattered within
the $n$th slice is given by

\begin{equation}
I^{scatt}_{ballistic}(n+1 \vert n) =
\exp\left(\frac{-ns}{l_{s}}\right) -
\exp\left(\frac{-(n+1)s}{l_{s}}\right).
\end{equation}

The calculation of the snake component is more involved and will be
discussed in later sections. However, if we denote $I_{trans}(n+1 \vert
n)$ as the sum of the unscattered and `snake' intensity that is
transmitted into the $n$th slice from the $(n-1)$th slice, then the
fraction of this intensity that undergoes scattering in travelling the
length of the $n$th slice, which is also the intensity of the source
$S_{n}$, is given by the Lambert-Beer law as

\begin{equation}
I^{scatt}_{total}(n+1 \vert n) = I_{trans}(n+1 \vert n)
\exp\left(\frac{-s}{l_{s}}\right)
\end{equation}

Of $I^{scatt}_{total}(n+1 \vert n)$, a fraction $\Gamma_{b}$ is
scattered into the rear hemisphere while a fraction $\Gamma_{f}$ is
scattered into the forward hemisphere. The fractions $\Gamma_{b}$ and
$\Gamma_{f}$ are calculated by numerically integrating the normalised
Mie scattering phase function, which is the angular scattering
probability density function, over the two hemispheres. Of interest
to us are the backscattered photons. These photons either travel
nearly undeviated and exit from the input face of the cell, or are
directionally randomised and converted to a diffuse flux along the
way. The former are what we have called the `reflected snake
photons'. These are photons that have travelled along the $z$ axis
in nearly straight-line paths and thus retained a significant
memory of their initial polarisation. After suffering a single
backscattering event they once again snake their way out of the
medium. A typical path is shown by the dotted line in Fig. 3. Since
the paths are significantly deviated only once, the reflected snake
intensity at the back face is in effect, almost the same as the
backscattered intensity from a dilute suspension of colloidal
particles where the photons scatter no more than once. Near the
exact backscattering direction, in which our observations are made,
these photons are responsible for the formation of the patterns.
Further away from the axis it would require a double scattering
sequence as described in \cite{Rakovic} for these photons to lie within
the acceptance angle of the detector. The forward scattered photons
may similarly, be carried over to the next slice or diffused. The
fraction carried over, we call the `carry-over' fraction. We make the
simplifying assumption that the carry-over fraction continues to
travel along the $z$ axis with the rest of the unscattered beam.
Thus, within each slice, photons may either be `carried-over' into the
next slice, converted into a diffuse flux or exit the medium as the
pattern forming reflected snake photons. The total fraction of the
incident intensity that exits in the form of reflected snake photons
is termed the `pattern forming' fraction.

The polarisation patterns are then obtained by mapping the Mie
scattering cross section onto the input face of the scattering cell as
follows. The scattered intensities are calculated over a 10 $mm^{2}$
area that is divided into a grid of 100 $\times$ 100 elements which we
refer to as pixels. For each pixel, we calculate the angle subtended
by the centre of the pixel at the source $S_{n}$ as shown in Fig. 3.
The flux passing through a given pixel, due to the intensity scattered
in the $n$th slice, is given by

\begin{equation}
I_{pixel} = I^{scatt}_{total}(n+1 \vert n) \Gamma_b
\sigma(\theta,\phi) \frac{dA}{2\pi r^{2}} {\bf P}(r)
\label{Pixel_int}
\end{equation}

It is to be remembered, that $I^{scatt}_{total}(n+1 \vert n)$ is the
sum of the fraction of the incident intensity that is scattered for
the first time, as well as the fraction of the carry-over flux from
{\it all preceding slices} that is scattered within the $n$th slice.
The fraction of this scattered flux passing through the pixel whose
centre makes an angle $(\theta, \phi)$ with the source, is given by
$\sigma(\theta,\phi) \frac{dA}{2\pi r^{2}}$, where
$\sigma(\theta,\phi)$ is the normalised Mie scattering cross section
at the angle $(\theta, \phi)$, and $\frac{dA}{2\pi r^{2}}$ is the
solid angle subtended by the pixel of area $dA$ at the source.
However, not all this flux is going to pass through the pixel since
multiple scattering converts a part of it to a diffusing flux. ${\bf
P}(r)$ is the probability that a photon will remain undeviated in
travelling a distance $r$. Consequently ($1-{\bf P}(r)$) is the
fraction of the flux converted to diffusive transport in traversing
the length $r$. Thus the net diffuse intensity in the $n$th slice,
which is a sum of the incident intensity that is diffused in
traversing the length $s$ of the slice, and the backscattered
intensity that is diffused as it travels towards 
a pixel at a distance $r$ on the input face of the
cell, is given by

\begin{equation}
I^{scatt}_{diffuse}(n+1 \vert n) = I^{scatt}_{total}(n+1 \vert n)
(1-{\bf P}(s)) + \sum_{all pixels}I_{pixel}(1-{\bf P}(r))
\end{equation}

The diffuse flux contributes a smooth randomly polarised background
with a $1/r$ variation \cite{Ping Sheng}, where $r$ is the distance
from the source of diffusing radiation to the pixel of interest. At
distances far from the point of entry of the beam, the diffuse
intensity dominates and no patterns are visible.

Mie scattering routines, including the one we have used \cite{Barber},
typically calculate the scattered fields along two orthogonal
directions ${\bf \hat e_{\bf \theta}}$ and ${\bf \hat e_{\bf \phi}}$,
which are the unit vectors of the spherical coordinate system. Given
the scattered fields $\left[ E_{\theta},E_{\phi} \right]$, the fields
$\left[ E_{x},E_{y} \right]$ in the lab frame, and the corresponding
intensities ($I = E \cdot E^{*}$) are obtained by the simple
polar-to-rectangular transformation

\begin{eqnarray}
{\bf E_{x}} = \cos\theta \cos\phi{\bf \hat e_{\theta}} - \sin\phi{\bf
		   \hat e_{\phi}}\\
{\bf E_{y}} = \sin\phi\cos\theta{\bf \hat e_{\theta}} - \cos\phi{\bf
		   \hat e_{\phi}} \label{matrix-trans}
\end{eqnarray}

Equations (1-4) completely determine the intensities of the sources
$S_{n}$. However, a crucial input to the calculation of the scattered
intensities, the function ${\bf P}(r)$ has not been defined, without
which $I^{scatt}_{total}(n+1 \vert n)$ cannot be calculated.
Diffusion theory maintains that photons are directionally randomised
on the scale of the transport mean free path $l^{*}$ and that the
probability for a photon to travel a path length $r$ without being
randomised is given by

\begin{equation}
{\bf P}(r)= \exp(-r/l^{*})
\label{diffpred}
\end{equation}

\begin{figure}[!h]
\centerline{\psfig{figure=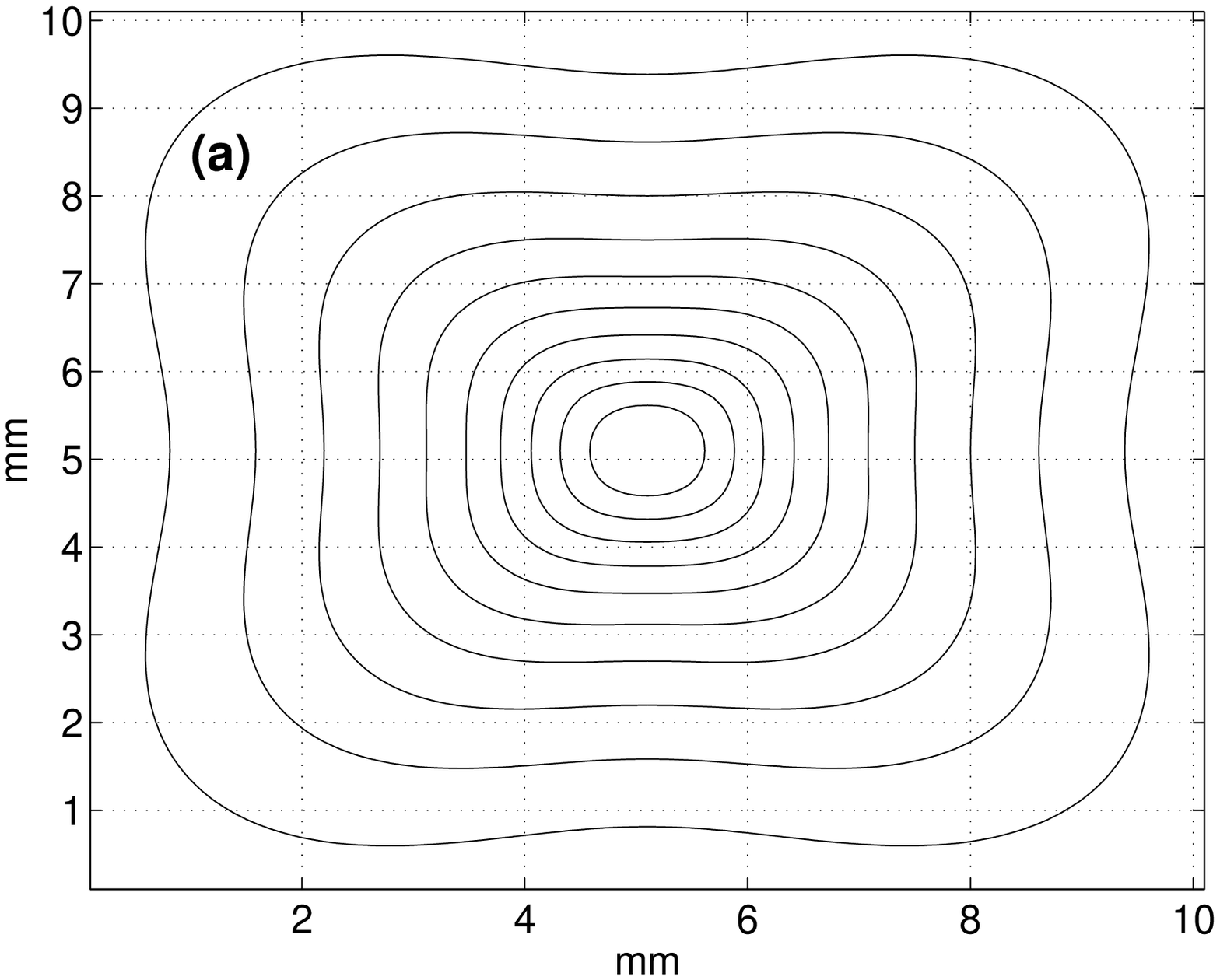,width=8cm,height=8cm}
\psfig{figure=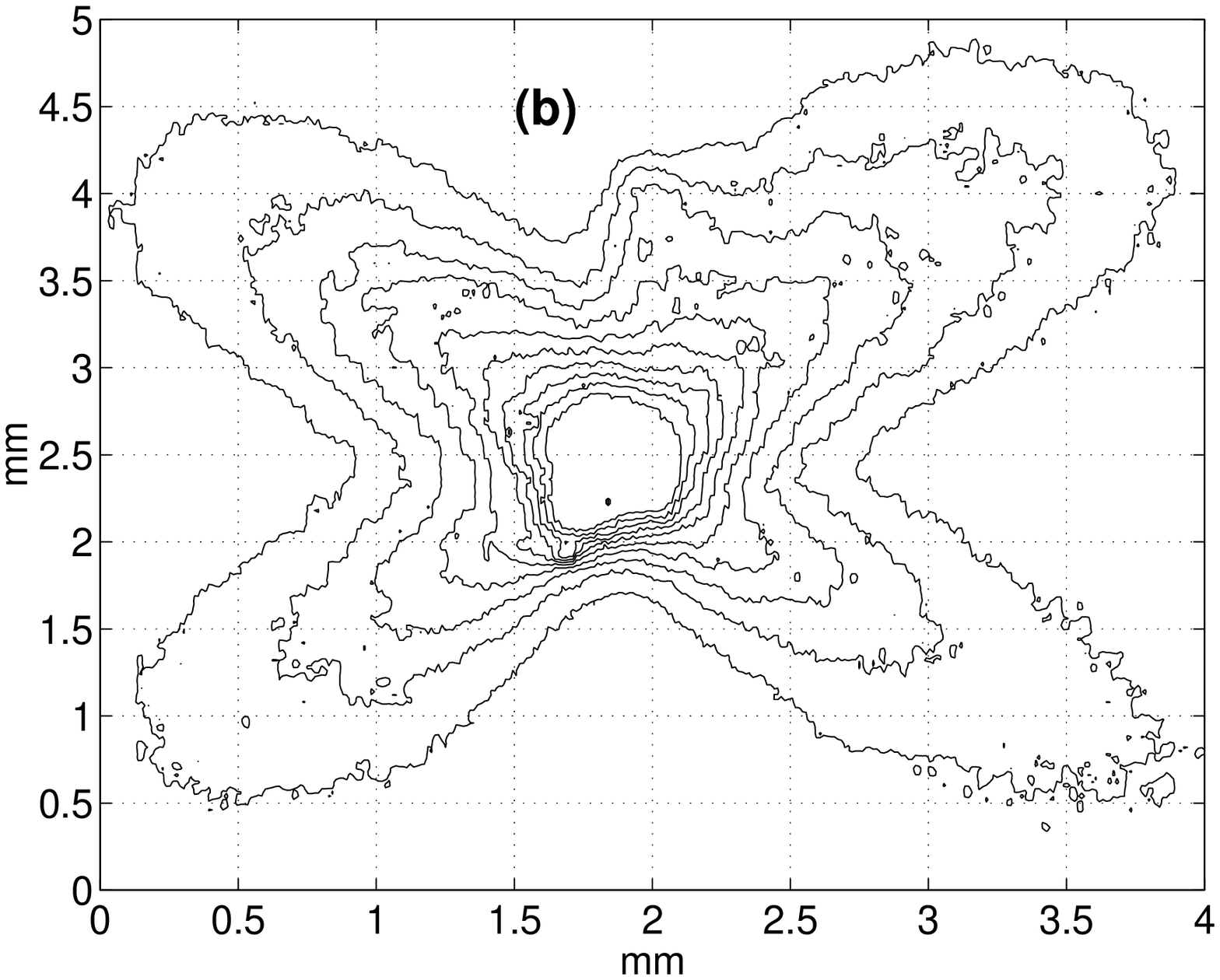,width=8cm,height=8cm}}
\caption{Calculated intensity contours obtained for a scattering suspension in
the VH geometry with $\tau = 0.5$ using the slice model are compared
with experimental results. Figure 4(a) is obtained assuming,
according to diffusion theory, that the probability that a photon will
travel a length $r$ without being randomised is given by $P(r) =
\exp(-r/l^{*})$. Figure 4(b) shows the experimentally obtained
contours at the same optical density. The figures bear almost no
resemblance to one another and is the motivation for the
re-examination of the assumptions underlying the diffusion
approximation.}
\end{figure}

and therefore, the snake fraction at a depth $r$ within the medium is
given by $I_{snake} = 1-{\bf P}(r)$. Figure 4(a) shows the pattern
generated for the backscattered intensity for a slab of thickness $L =
0.5l^{*}$, assuming the form for ${\bf P}(r)$ given by diffusion
theory. Figure 4(b) shows the experimentally obtained pattern at the
same optical density. We find that the patterns bear little
resemblance to one another because the diffuse flux, as given by eq.
(\ref{diffpred}), is exceedingly large even at moderate optical
densities. The diffuse intensity may be lowered if we allow the
production of diffuse photons to take place on a length scale much
larger than $l^{*}$ and hence much slower than the exponential decay
predicted by eq.(\ref{diffpred}). Consequently, it also implies that
the snake photons must travel far deeper into the medium than assumed
by diffusion theory, a result that we have reported and substantiated
with experimental data in \cite{Venkatesh2}. To understand these issues
better, we have performed Monte Carlo simulations, which are described
in the next section.

\section{Monte Carlo simulations}

The procedure for our Monte Carlo simulations was as follows. Photon
transport within a slab was modelled assuming that the photons
travelled exponentially distributed lengths $s$ between scattering
events. The probability $P(s)$ of travelling a ballistic path length
$s$ is given by the familiar Lambert-Beer law, $P(s) = exp(-s/l_s)$,
where $l_s$ is the scattering mean free path of the photons in the
medium. The random paths between scattering events were generated
taking $s = -l_s \cdot \ln(RAN)$, where RAN is a random number
uniformly distributed between 0 and 1 \cite{Numrec}. The scattering
angles were chosen such that they had a distribution of directions
given by the Henyey-Greenstein phase function \cite {Durian1}, where
the probability of scattering at an angle $\theta$ relative to the
incident direction of the photon is given by 

\begin{equation}
P(\cos \theta) = \frac{1 - g^2}{(1 + g^2 - 2g \cos \theta)^{3/2}}
\end{equation}

\subsection{Defining a `diffuse' photon?}

\begin{figure}[!h]
\centerline{\psfig{figure=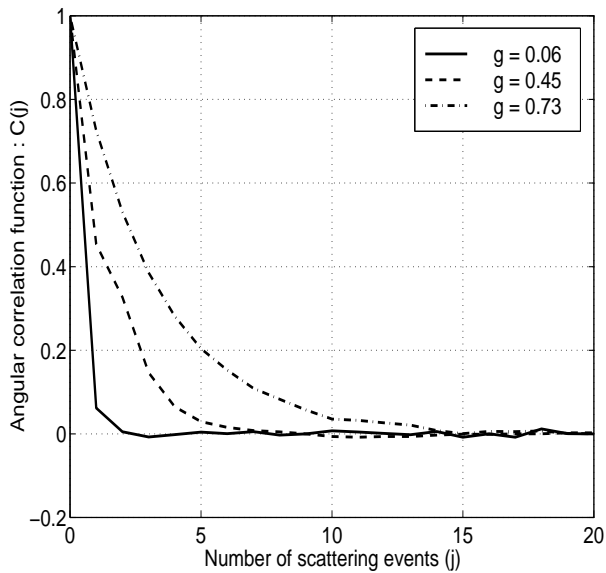,width=8cm,height=8cm}}
\caption{The directional correlation function $C(j) = \langle {\bf \hat{n}}(0)
\cdot {\bf \hat{n}}(j) \rangle$ is shown for three different values of
the scattering anisotropy, $g = 0.06, 0.45$ and $0.73$}
\end{figure}
In order to understand the process of conversion of the incident
ballistic photons to a diffusive flux, we need an operational
definition of a diffusing photon. In the spirit of diffusion theory,
we assume that a diffuse photon is one which has lost all memory of
its initial direction of propagation. The scattering anisotropy is a
measure of the ability of a photon to `remember' its initial
direction. For a scattering anisotropy $g = 1$, the photon always
travels along its initial direction of propagation, while for $g = 0$,
it is randomised at each scattering event. We used random walk
simulations to investigate, as a function of the scattering
anisotropy, the number of scattering events that it takes for a photon
to be completely randomised directionally. Photon trajectories were
generated, as described previously, in an infinite medium. The
directional correlation function $C(j) = \langle {\bf \hat{n}}(0)
\cdot {\bf \hat{n}}(j) \rangle$, where ${\bf \hat{n}}(j)$ represents
the unit vector along the photon trajectory after the $j$th scattering
event, was calculated for eight different values of $g$ by averaging
over $10^{4}$ trajectories. These results are shown in Fig. 5. At
long times, $C(j)$ fluctuates about zero. We calculate the standard
deviation in the fluctuations of $C(j)$ at long times and we find the
point when $C(j)$ drops below this value for the first time. This is
taken as the number of scattering events $N_{d}$, required for a
complete loss of angular correlation. It can be seen from Fig. 5
that the conversion to diffusive motion is more rapid in the case of
the isotropic scatterers as it takes only a few, ($N_{d} \sim 3$),
events for angular memory to be lost. On the other hand, the
anisotropic scatterers typically lose angular memory over many more
scattering events ($N_{d} \sim 9$ for $g = 0.423$ and $N_{d} \sim 17$
for $g = 0.732$). However, $N_{d}$ contains no information on the
length scale over which this randomisation of the direction proceeds.
The next question therefore that we need to answer is how far away
from the source are the photons after undergoing $N_{d}$ scattering
events?. This yields the probability ${\bf P}(r)$ that is central to
the slice model, and is discussed in the next subsection.

\subsection{Penetration depth of the `snake' photons}

According to diffusion theory, the randomisation of a photon takes
place typically after travelling a distance $l^{*}$, and usually the
source of diffusing photons is modelled as a delta function at a depth
of $l^{*}$ within the medium. However, it is also well known that
this assumption, especially when considering photon transmission, is
inaccurate while the thickness of the scattering medium is less than
about $8l^{*}$ \cite{Kaplan}. Clearly, photons are not randomised at
any one single depth inside the medium and obviously there exists a
smooth distribution of lengths over which the initially
quasi-ballistic flux is converted to a diffusive one.

\begin{figure}[!ht]
\centerline{\psfig{figure=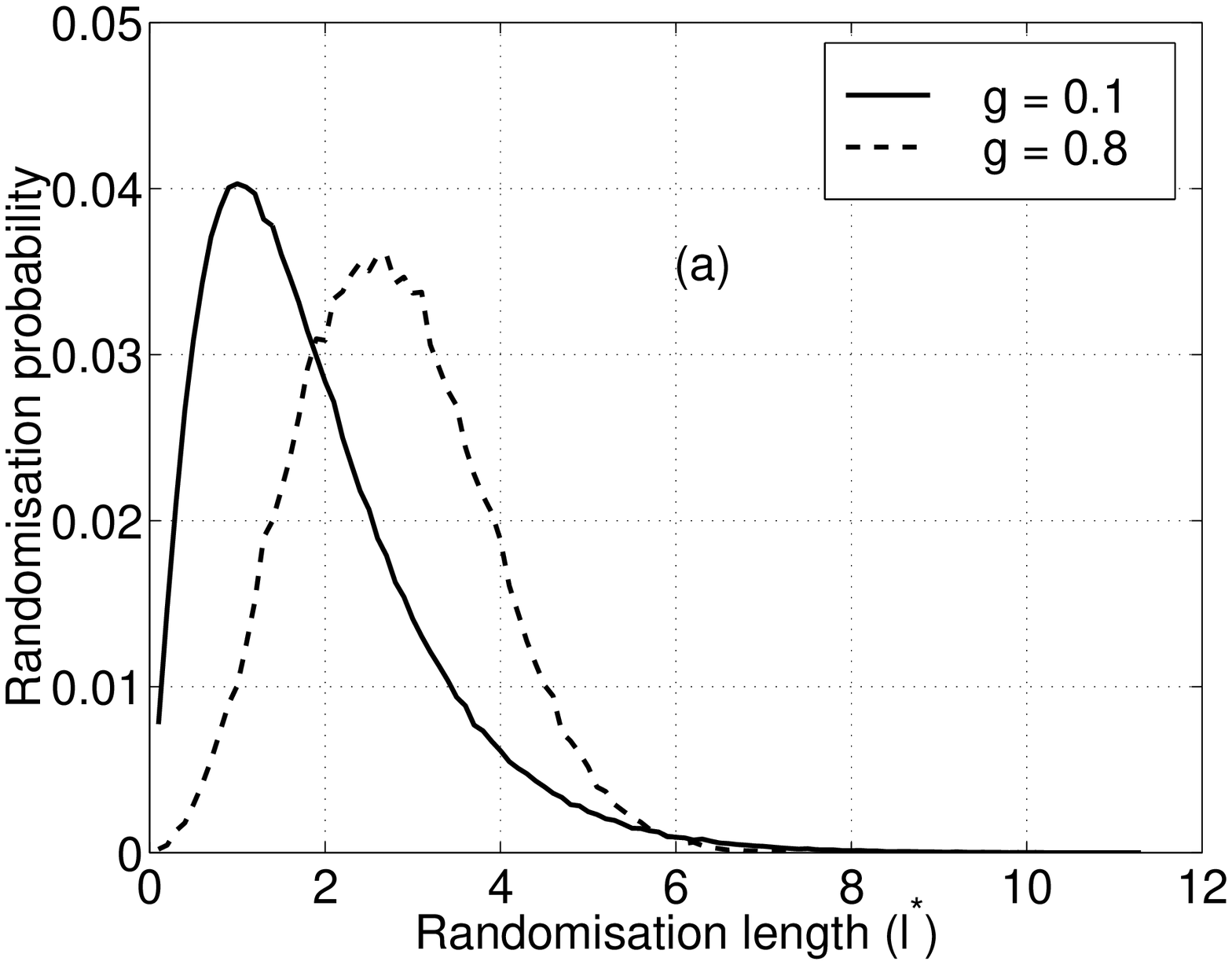,width=8cm,height=8cm}
\psfig{figure=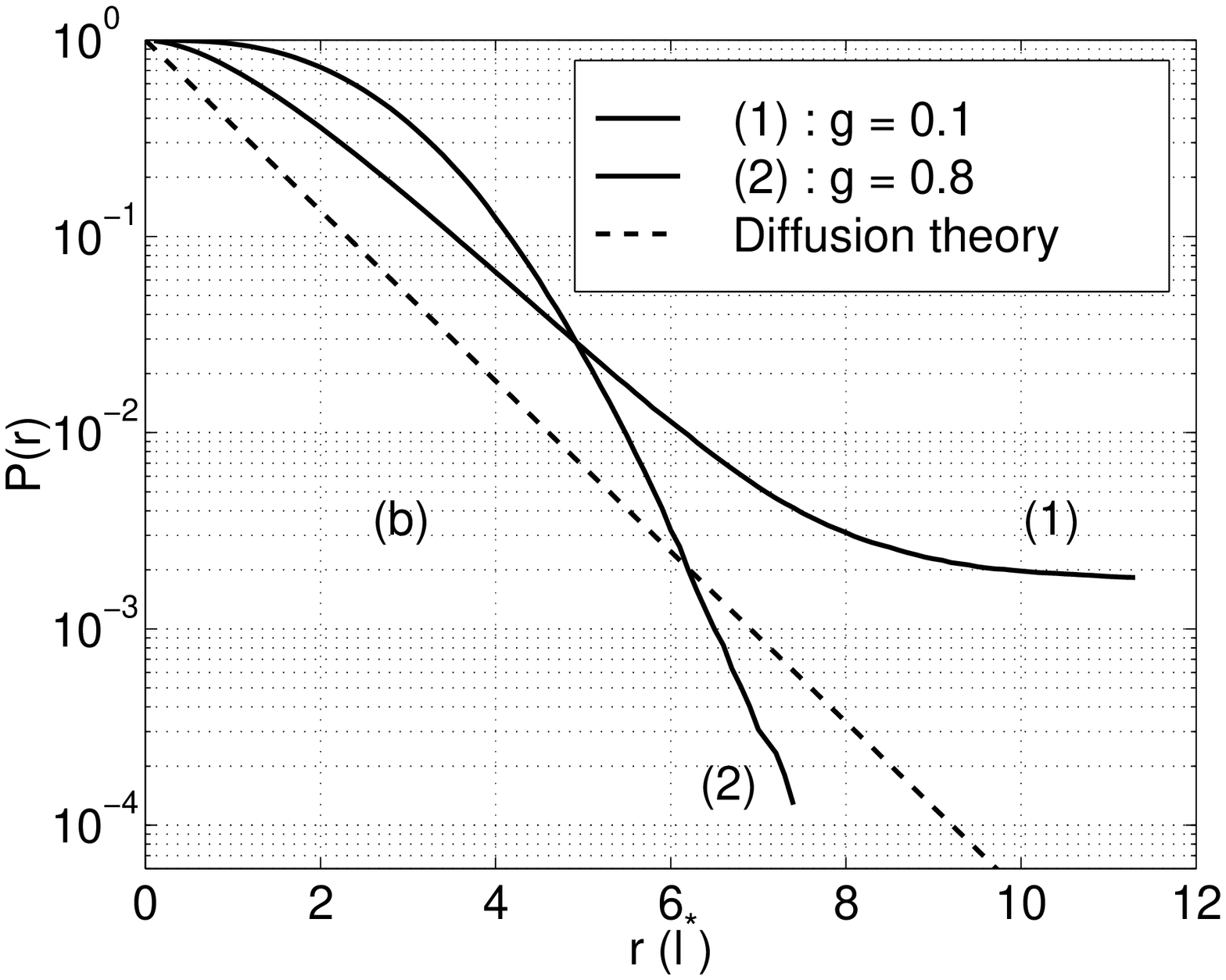,width=8cm,height=8cm}}
\caption{Understanding the penetration depth of the snake photons :  Figure
6(a) shows the normalised histograms obtained for the randomisation
probability that a photon will undergo $N_{d}$ scattering events as a
function of the distance travelled from the source. This distance is
termed the `randomisation length'. The $x$ axis is scaled by the
transport mean free path $l^{*}$. Integrating the randomisation
probabilities yeilds the cumulative probability $1-{\bf P}(r)$ that a
photon will be randomised in travelling a distance $r$ into the
medium, while ${\bf P}(r)$ is the probability that a photon will
remain undeviated on travelling a length $r$. Figure 6(b) shows a
semilogarithmic plot of the probability ${\bf P}(r)$ that is used to
calculate the source intensities in the slice model. The
randomisation probabilities and the corresponding ${\bf P}(r)$ are
show for $g = 0.1 and 0.8$.}
\end{figure}
Random walk simulations were used to obtain this distribution of
lengths. Once again photons were launched from the origin, but in a
semi-infinite half-space instead. For a given scattering anisotropy
though, we know from the previous simulation, the number of scattering
events $N_{d}$ required for the photon to lose directional memory.
Thus, the simulation propagated photons as before but terminated the
trajectory either when the photon had scattered $N_{d}$ times, or if
the photon had been backscattered out of the half space. Boundary
reflections were neglected and absorbing boundary conditions were
applied at the back face of the semi-infinite medium. The radial
distance travelled from the origin after $N_{d}$ scattering events was
stored and a histogram of the distribution of these radial distances
was constructed. Simulations were carried out for the following
anisotropy values :  $g = $ 0.1, 0.2, 0.3, 0.4, 0.5, 0.6, 0.8 and 0.9.
The histogram bins had a width of $l^{*}/10$ and were normalised by
the total number of photons that underwent $N_{d}$ scattering events.
In the subsequent discussion, after this normalisation, we treat the
number stored in each bin as the randomisation probability. This
however is strictly not true and a detailed discussion of the
consequences of our choice of the normalisation may be found in
\cite{Venkatesh2}. Figure 6(a) shows a plot of the histograms
obtained for $g =$ 0.1 and 0.8. Integrating the area under the curves
yields the cumulative probability $1-{\bf P}(r)$ that a photon will be
randomised on travelling a distance $r$. Figure 6(b) shows a
semilogarithmic plot of the function ${\bf P}(r)$, the probability
that a photon will remain almost undeviated on travelling a distance
$r$, for $g =$ 0.1 and 0.8. The dashed line is the diffusion theory
prediction for ${\bf P}(r) = \exp(-r/l^{*})$, and it is evident that
the diffusion approximation greatly underestimates the snake photon
intensity.

\subsection{The offset length for the source of diffusing photons}

Given the probability ${\bf P}(r)$, we can calculate the fraction of
the incident flux that is converted to diffusive transport within a
distance $r$ from the point of entry into the medium. In the slice
model, all the intensity scattered within one slice length, is assumed
to radiate from a source at the centre of the next slice. However, we
found that placing the source of diffuse radiation in this manner {\it
still} overestimated the diffuse intensity which continued to dominate
the backscattered intensity, though not as strongly as before.
Clearly, it is not only the diffuse intensity predicted by the
diffusion approximation that needs to be reexamined, but also the
apparent position of the source of this intensity. What then must we
consider as the source of diffusing radiation?. By definition, the
diffuse radiation is emitted equally in all directions. Therefore,
that point about which the diffuse flux is spherically symmetric must
be chosen as the source of the diffusing photons. In other words,
given a cloud of diffusing photons, the apparent source lies at the
`centre of mass' of the cloud.
\begin{figure}[!h]
\centerline{\psfig{figure=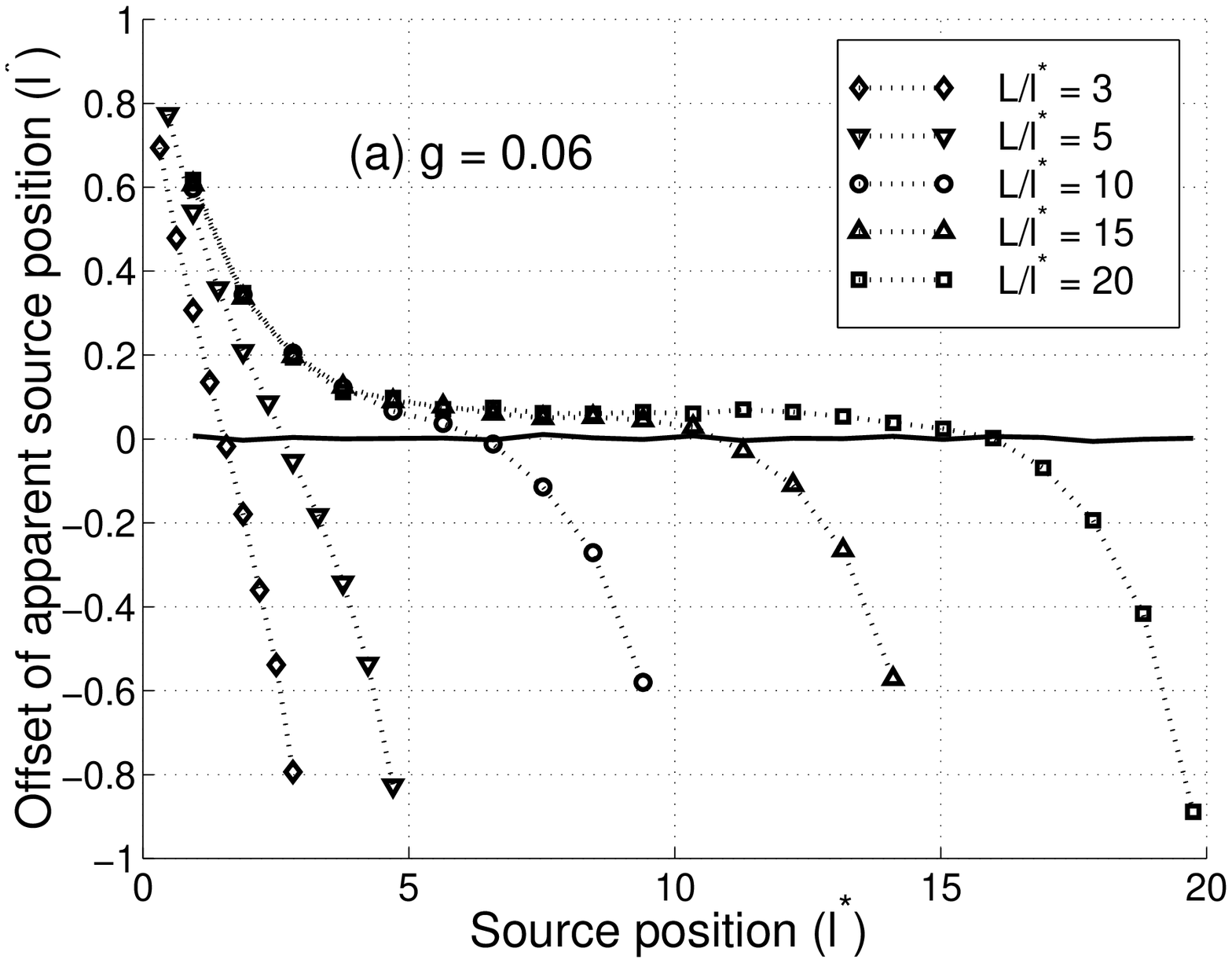,width=8cm,height=8cm}
\psfig{figure=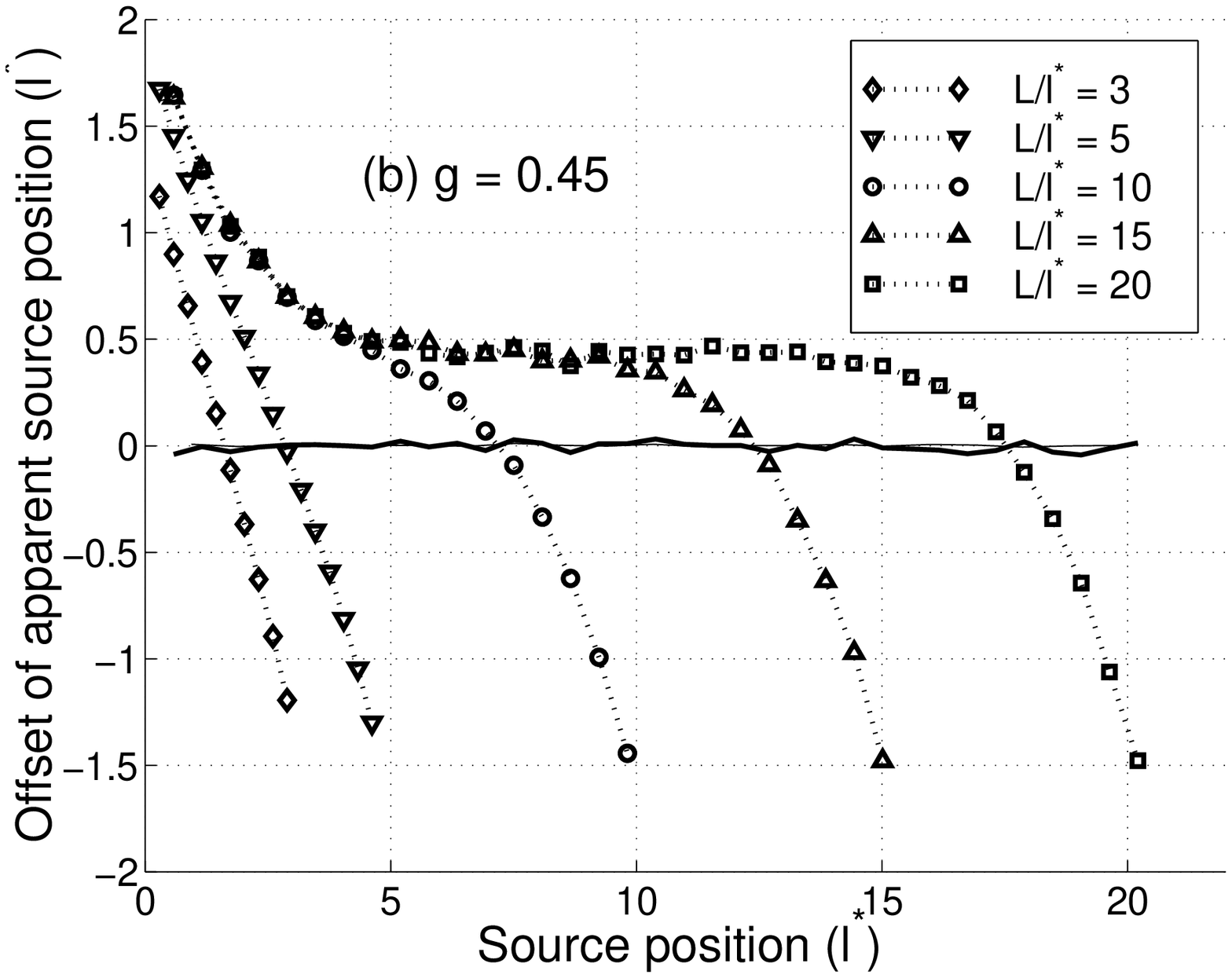,width=8cm,height=8cm}}
\centerline{\psfig{figure=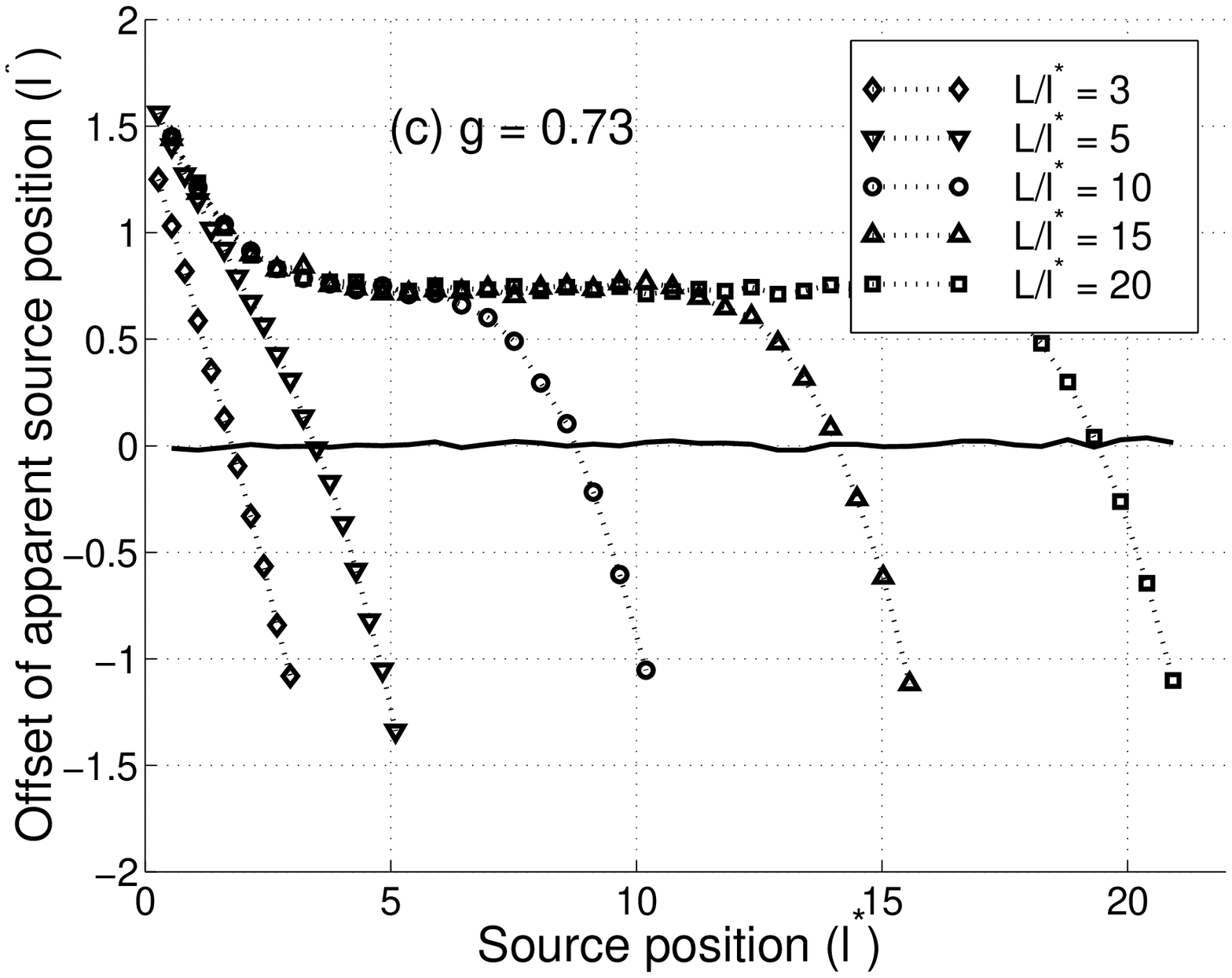,width=8cm,height=8cm}}
\caption{The offset length is the difference in the $z$ coordinates of the
centre of mass of the diffuse photon cloud after all the photons have
undergone $N_{d}$ scattering events, and the point at which these
photons were launched. The figures show the variation in the offset
length for different thicknesses $L$, of the scattering medium, and
also for different values of the scattering anisotropy $g$. The
figures (a), (b) and (c) correspond to $g$ values of 0.06, 0.452 and
0.732 respectively.}
\end{figure}
We calculated the position of the centre of mass by modifying our
earlier simulation. Photons were launched at a number of sources
$S_{n}$ as in the slice model, within a box of thickness $L$ and
infinite transverse extent. Once again photons were propagated until
they underwent $N_{d}$ scattering events after which the trajectories
were terminated as before and their coordinates stored in an array.
The mean coordinates of the photon cloud, $<x>, <y>$ and $<z>$, were
computed by averaging over the coordinates of all photons after they
had undergone $N_{d}$ scattering events. The simulations were
performed for 5 different box thicknesses to study the effect of the
boundaries upon the position of the apparent source of diffusing
photons. Figures 7(a), 7(b) and 7(c) show the results obtained for
three different scattering anisotropies.

On the $x$ axis, we have in units of $l^{*}$, the position of the
source of photons with relation to the ends of the box, with 0 being
the left extreme of the box. The mean value of the $z$ coordinate of
the cloud of diffusing photons is plotted on the $y$ axis. A positive
value indicates that after $N_{d}$ scattering events, the centre of
mass of the photon cloud lies ahead of the point at which the photons
were launched while a negative value indicates that it lies behind the
launch position (Assuming that the photons enter the medium at the
left and move along the $+z$ direction). The length between the
centre of mass or apparent source and the real source is called the
offset length ${\bf L_{offset}}$. The solid line along the $x$ axis
at $y = 0$, is a plot of $<x>$ and $<y>$. It shows that the photon
cloud is symmetric about the $z$ axis along which the beam enters the
medium, which is as one would expect, for all azimuthal directions are
equivalent in our model. Common to all the curves for all values of
$g$ is the fact that when the source of photons lies close to the
walls, the centre of mass, or the apparent diffuse source lies deeper
within the medium than does the source. For photons entering the
medium from the $z = 0$, the apparent source of diffuse radiation lies at
a distance $l^{*}$ or greater within the medium.

On comparing Figs. 7(a), 7(b) and 7(c), two features are noteworthy.
The first is that we may estimate from the variation of the offset
length with source position, the penetration depth of the snake
photons. The offset length shows an initial exponential decay as the
source is moved farther into the medium after which it remains flat
for a while and decays sharply once more. The latter part of the
curve appears nearly like an inversion about the $x$ axis, of the
initial exponential decay. The point at which the curve begins to
stay flat is the depth beyond which the photons do not feel the
influence of the boundaries any longer. When the photon cloud is
symmetric, the presence of the boundary does not affect the offset
length. Therfore, as long as the offset is influenced by the
boundaries, it indicates an asymmetry of the photon cloud, which
exists only when there are photons which still retain some memory of
their initial direction of propagation. Thus we may infer indirectly
the penetration depth of the snake photons. It is known
experimentally \cite{Kaplan, Pine} that the minimum thickness
$L_{min}$ for which the diffusion approximation yields accurate
results in the analysis of multiple scattering experiments, is of the
order of 6-10$l^{*}$. It has also been observed experimentally that
$L_{min}$ is inversely related to the scattering anisotropy $g$
\cite{Kaplan}. Our results reproduce the trends observed by Kaplan
{\it et.al}. At $g = 0.06$, the curve flattens at a depth of
approximately 8$l^{*}$, while for $g = 0.73$, the depth is reduced to
about 4$l^{*}$. We have reported a similar result in
\cite{Venkatesh2}. The second feature that is to be noted is that far
inside the medium, when the sources no longer are affected by the
boundaries, the apparent source {\it always} lies ahead of the real
source by a length $gl^{*}$, {\it i.e.}  ${\bf{L}_{offset}} = gl^{*}$.
To our knowledge, this observation has not been reported in the
literature so far.

Thus in the slice model, we calculate the offset length for
each slice and the source of diffusing radiation is placed at the
position $(x = 0, y = 0, z = z_{slice} + $offset, where $z_{slice}$ is
the $z$ coordinate of the slice.

\section{The slice model revisited}

Incorporating the results of the simulations, we can now modify the
slice model to obtain the backscattered intensity patterns. In brief,
the two main results of our simulations have been, obtaining the
probability ${\bf P(r)}$ that a photon will travel a distance $r$
without being randomised, and the offset length of the apparent source
of diffuse photons. The computation of the source intensities is
performed in two steps. On the first pass, using ${\bf P}(r)$ from
the simulations and equations (1-6), our program calculates the
diffuse and carry-over intensities at each slice. On the second pass,
the carry-over intensity from each slice is propagated along the
length of the scattering medium and the contributions to the image
forming and diffuse intensities at each slice are calculated. The
azimuthal patterns are then formed by calculating the pixel
intensities on the back face from Eq. (\ref{Pixel_int}). Finally,
the diffuse intensity with a $\frac{1}{r}$ variation is then added to
each pixel. Since the source of diffusing radiation must be displaced
from the source $S_{n}$ by the offset length, the length $r$ is now
the distance between the apparent source at a position $S_{n} + {\bf
L_{offset}}$, and the pixel of interest. In calculating the diffuse
backscattered intensity, one additional assumption is involved. Since
the diffusing photons form a spherically symmetric cloud, we assume
that the fraction of the diffuse intensity that scatters into the back
face is proportional to the solid angle subtended by the back face at
the apparent source of diffuse radiation, an approximation that has
been used with good results previously \cite{Prasad}.

\begin{figure}[!h]
\centerline{\psfig{figure=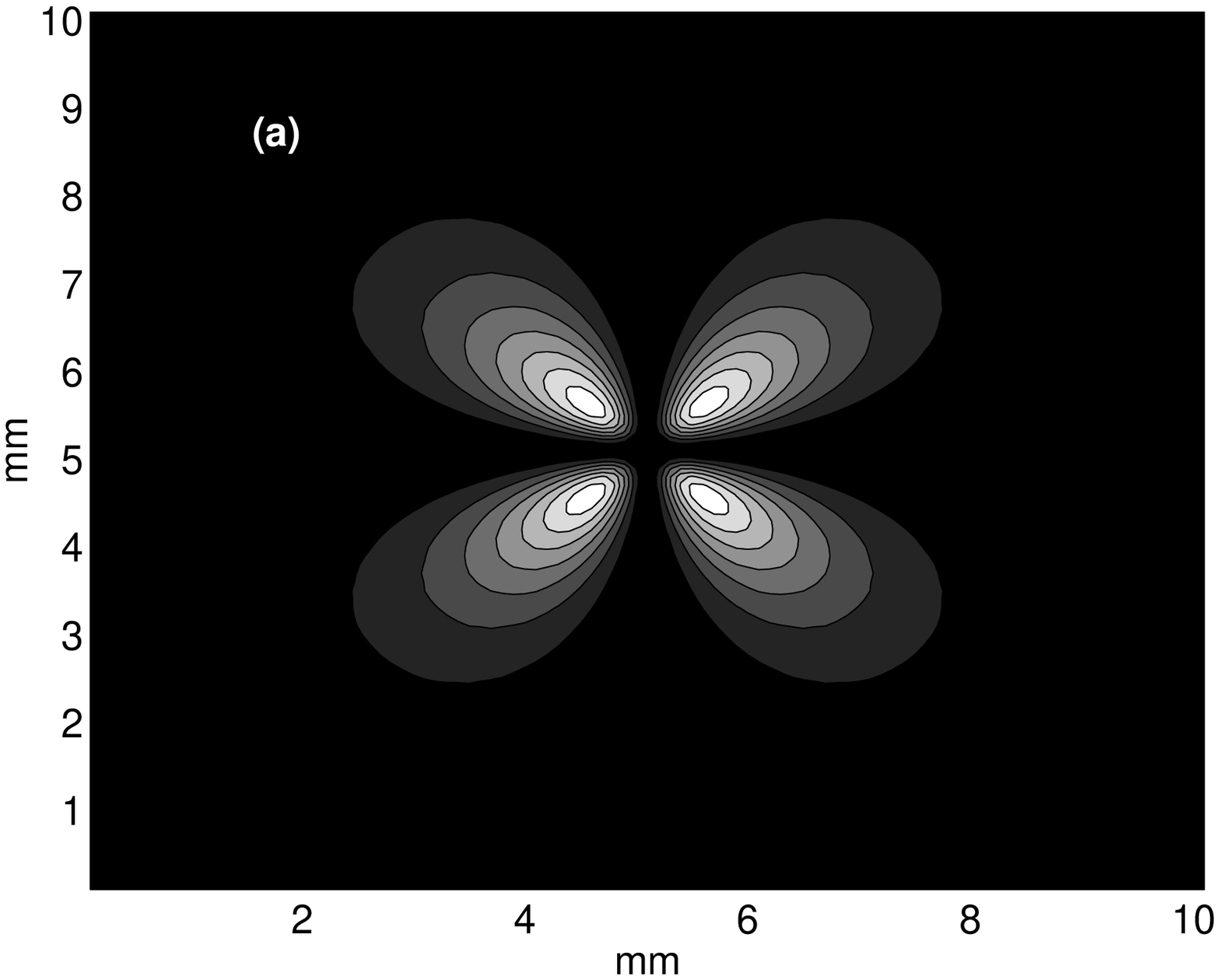,width=8cm,height=8cm}
\psfig{figure=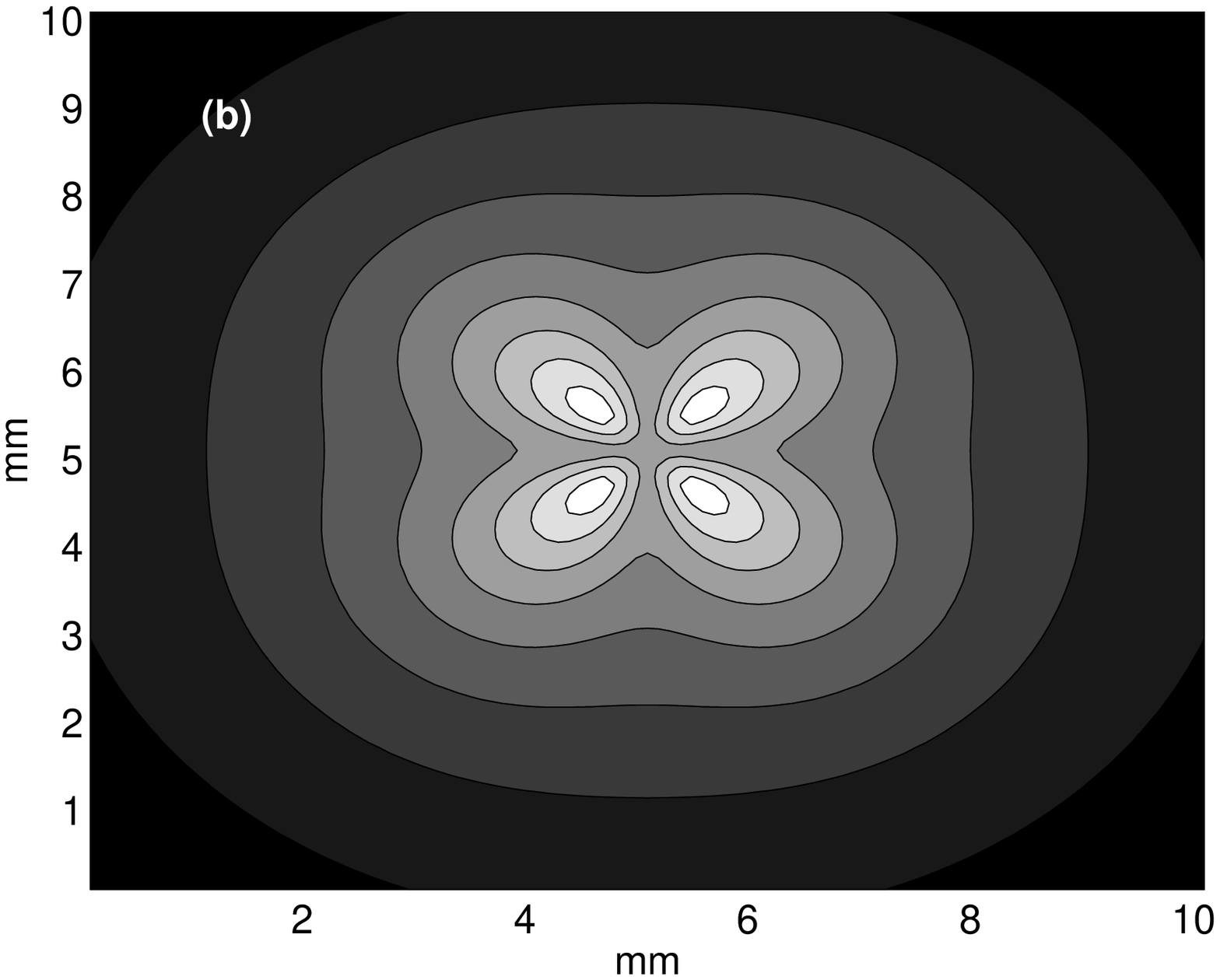,width=8cm,height=8cm}}
\caption{Results obtained using the slice model to calculate the backscattered
intensity patterns for a slab of optical density $\tau = 8.76$, in the
VH scattering geometry. Figure 8(a) computes the scattering without
including the diffuse scattered intensity, while the diffuse intensity
is also computed in Fig. 8(b).}
\end{figure}
However, before we can compare experiment with the results of our
model, another surprise awaits us. Figure 8(a) shows the calculated
backscattered intensity due to the reflected snake photons,for a slab
with $\tau = 8.76$, in the VH geometry. We see that along the $x$ and
$y$ axes the intensity is a minimum and no depolarised photons are
present. One obvious source of depolarised light is the diffuse flux.
In Fig. 9(b), the diffuse flux has also been incorporated with the
the source being placed at the correct offset length. Comparison with
experiment showed that the model accurately described the intensity
contours far from the centre, but contrary to experimental fact, the
central region is still seen to have a lower intensity than the
surroundings. Comparing these results with our experimental data
shown in Fig. 2, we observe that the central region is {\it always}
the brightest region in the image. Even, when the diffuse flux is
negligible, as in the case when $\tau = 0.51$ (Fig. 4(b)), we find
the well defined circular bright region at the centre of the image.
This immediately eliminates the diffuse flux as the cause for the
central bright spot, leading us to look for a scattering process that
scatters strongly around the exact backscattering direction, but which
falls off rapidly as we move away.

\subsection{Depolarisation}

What then is the source of this discrepancy?  We now argue that it is
due to the incoming intensity that is depolarised. The incoming light
is depolarised with each scattering event. However, depolarisation
and the length scale on which it proceeds, is currently a poorly
understood phenomenon. For our purposes, we require a model that
predicts how much of the snake component is depolarised as it
propagates across each slice. In the absence of such a model, we have
assumed that a small fraction of the total backscattered intensity
$I^{scatt}_{total}(n \vert n-1) \Gamma_b$ is depolarised at each
slice. The scattering phase function for a randomly polarised
incident intensity has just the properties that we are seeking. It is
sharply peaked near the exact backscattering direction and, when
incorporated in the slice model, reproduces accurately all the
experimentally observed features. We have found that a depolarised
source which scatters 2\% of the total backscattered intensity at each
slice gives very good agreement with experimental results.

\subsection{Comparison with experiment} 

Contour plots comparing experimental data with the computed intensity
contours are shown in Figs. 9 and 10. Contours are shown for the PO
and VH geometries. The distortion in the experimental contours is
because the input face of the sample cell are is not exactly normal to
the incident beam so that light reflected from the glass walls of the
sample cell is not scattered back into the CCD camera.

\begin{figure}[!h]
\centerline{\psfig{figure=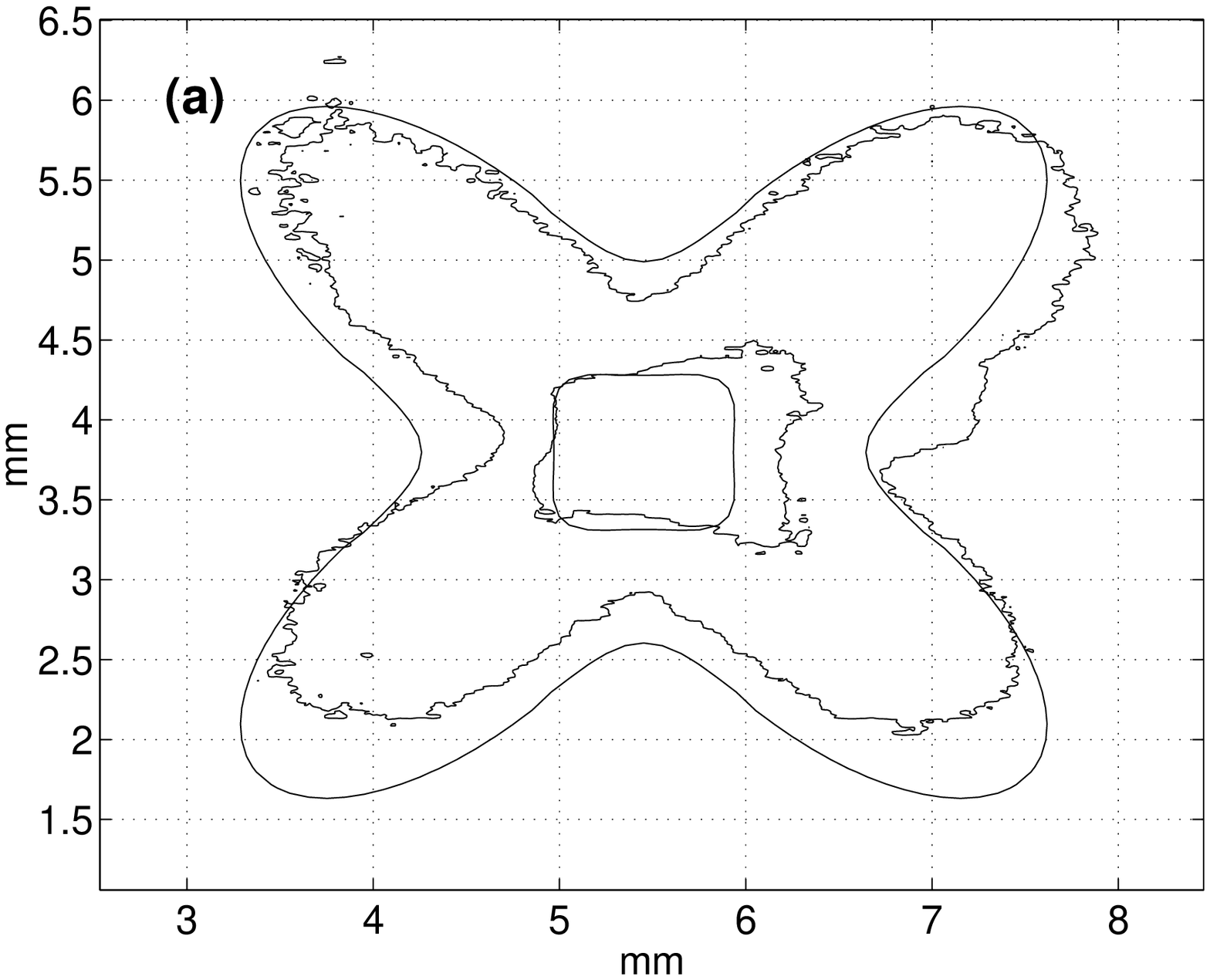,width=8cm,height=8cm}
\psfig{figure=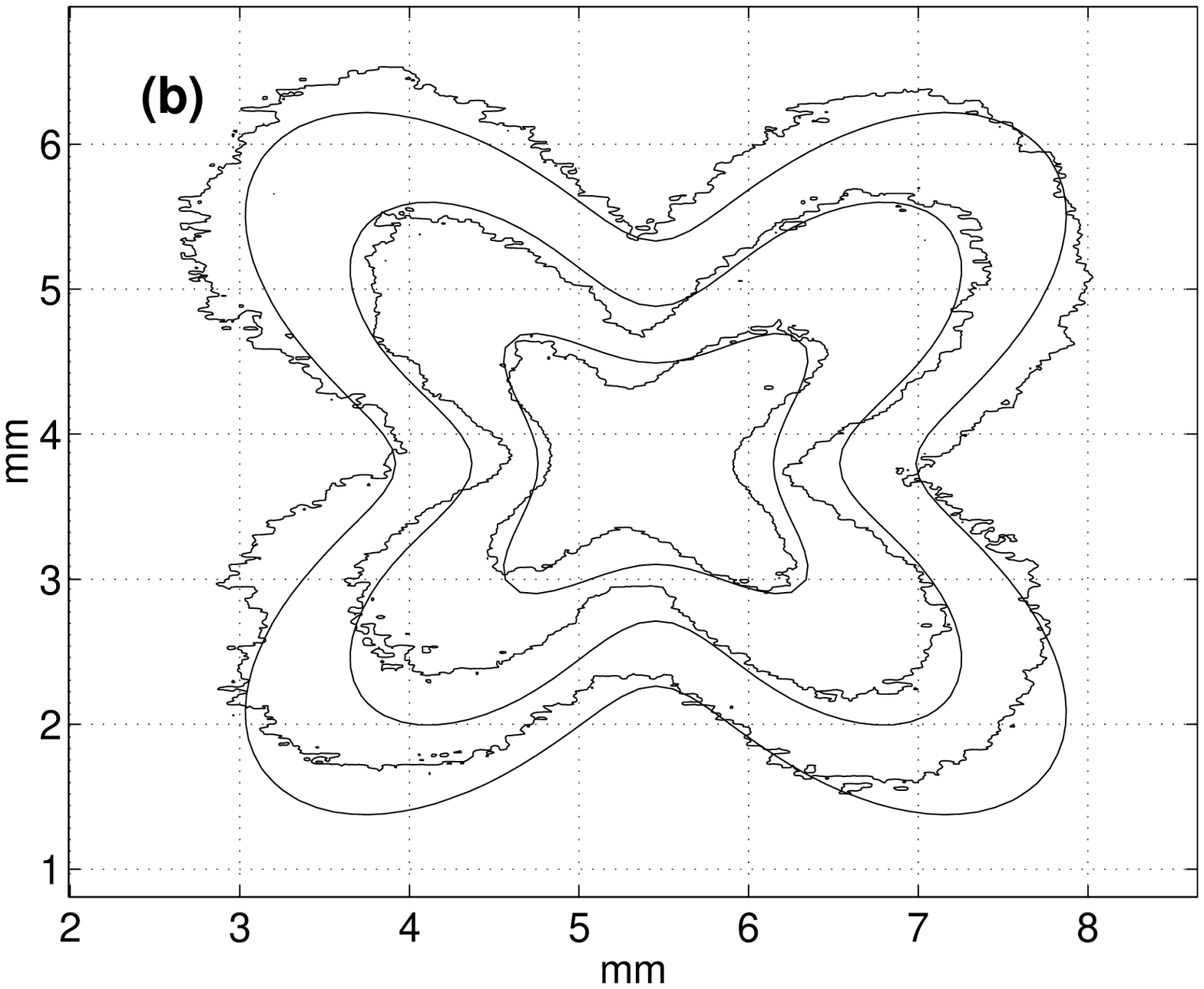,width=8cm,height=8cm}}
\centerline{\psfig{figure=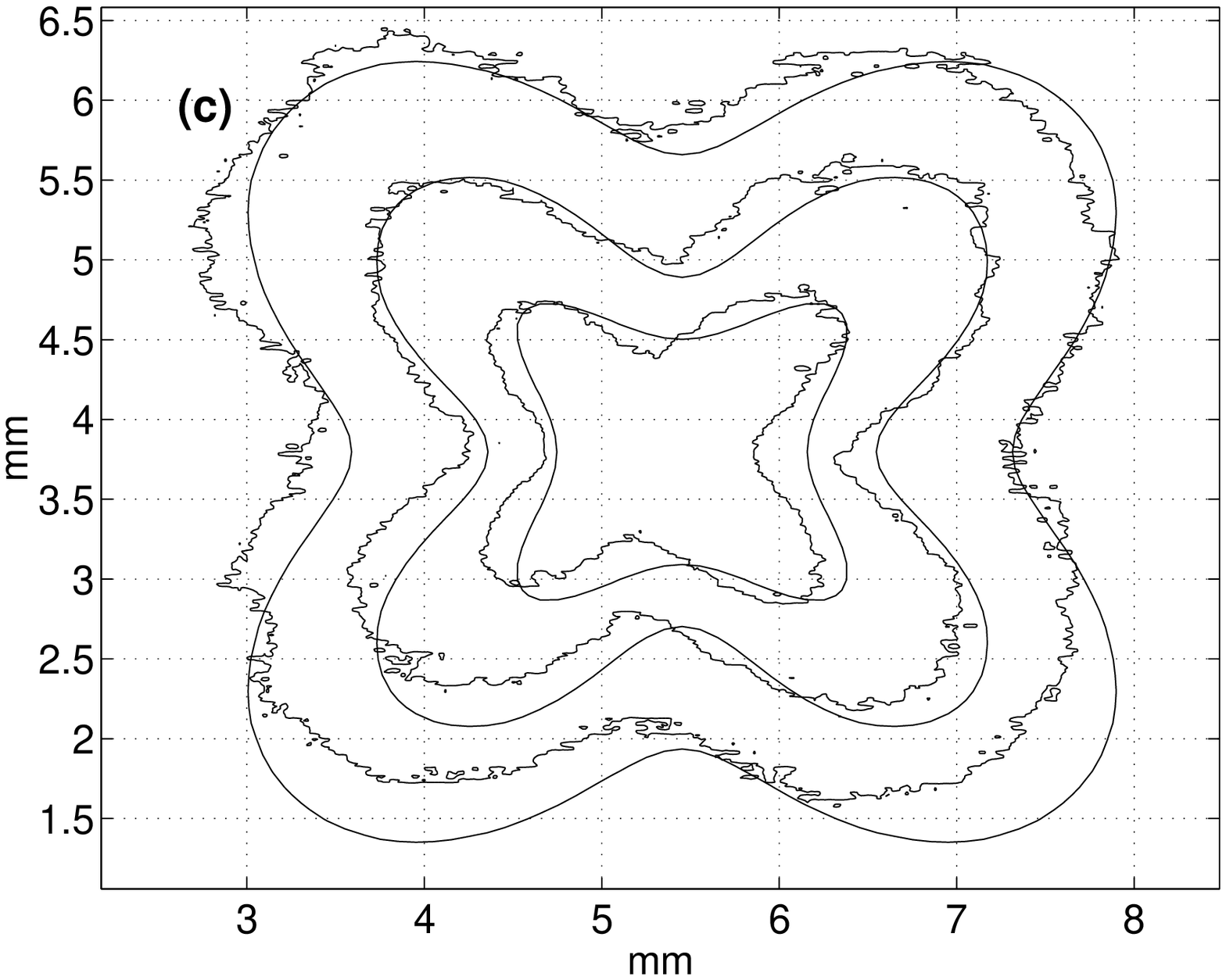,width=8cm,height=8cm}
\psfig{figure=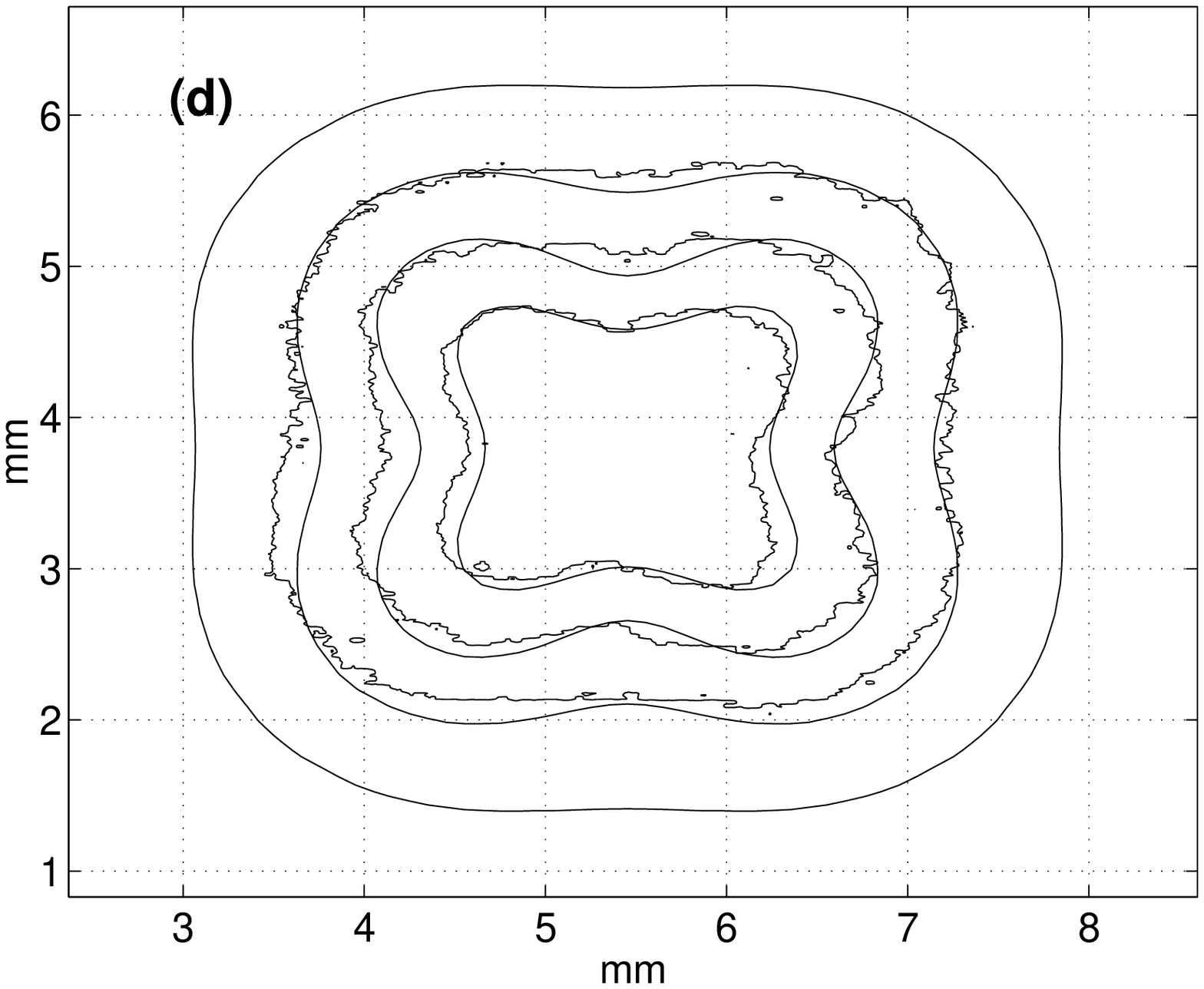,width=8cm,height=8cm}}
\caption{Comparison of experimentally obtained backscattering intensity
contours with the contours calculated by the slice model. The
scattering is in the VH geometry. The smooth lines are the calculated
contours while the jagged lines are experimental data. The figures
represent optical densities of $\tau$ = (a) 0.51, (b) 1.84, (c) 3.56,
and (d) 8.7. respectively.}
\end{figure}
Here we wish to make a comparison with the data of Heilscher {\it et
al} \cite{Bigio1}, who have obtained very high quality images with a
much larger dynamic range than ours by using a 14 bit CCD camera. In
their experiment, they place a circular mask at the centre of the
image which rejects specular scattering and also, almost completely,
the depolarised scattered light whcih causes the central bright spot.
In Fig. 3(a) of their paper, the backscattered intensity pattern
obtained in the VH geometry shows all the features present in Fig.
8(b). The four bright spots symmetrically distributed around the
central region are clearly seen. This further validates our
confidence in having captured the essential physics with the slice
model. We chose not to place a similar mask in our images since our
CCD viewing area was less than half the area in their experiments. To
mask the central bright spot completely requires a mask of about 1.5
to 2.0 mm diameter. This not only greatly reduces our image area, but
the mask also distorts the image in its immediate vicinity, further
reducing an already small viewing area.

\begin{figure}[!h]
\centerline{\psfig{figure=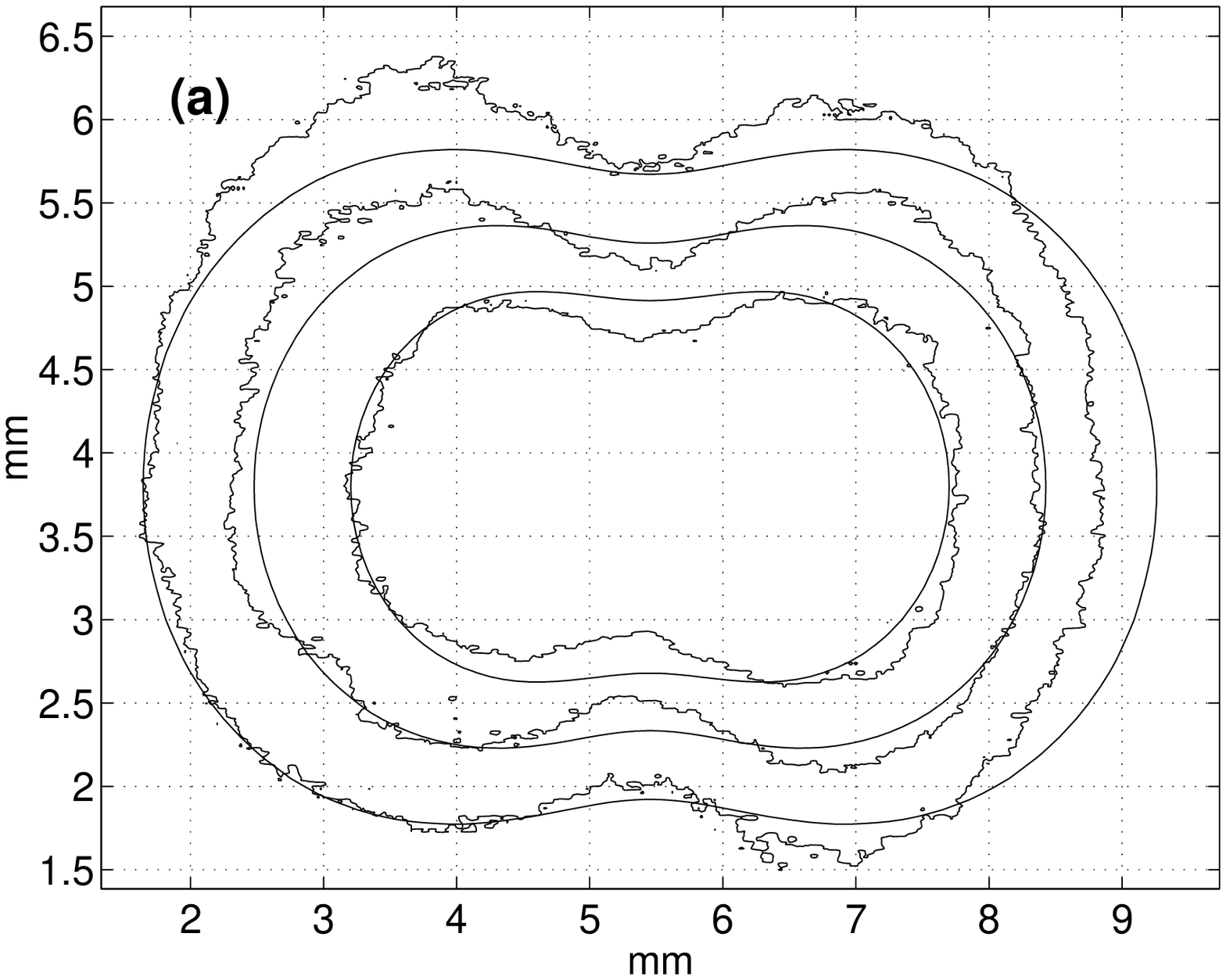,width=8cm,height=8cm}
\psfig{figure=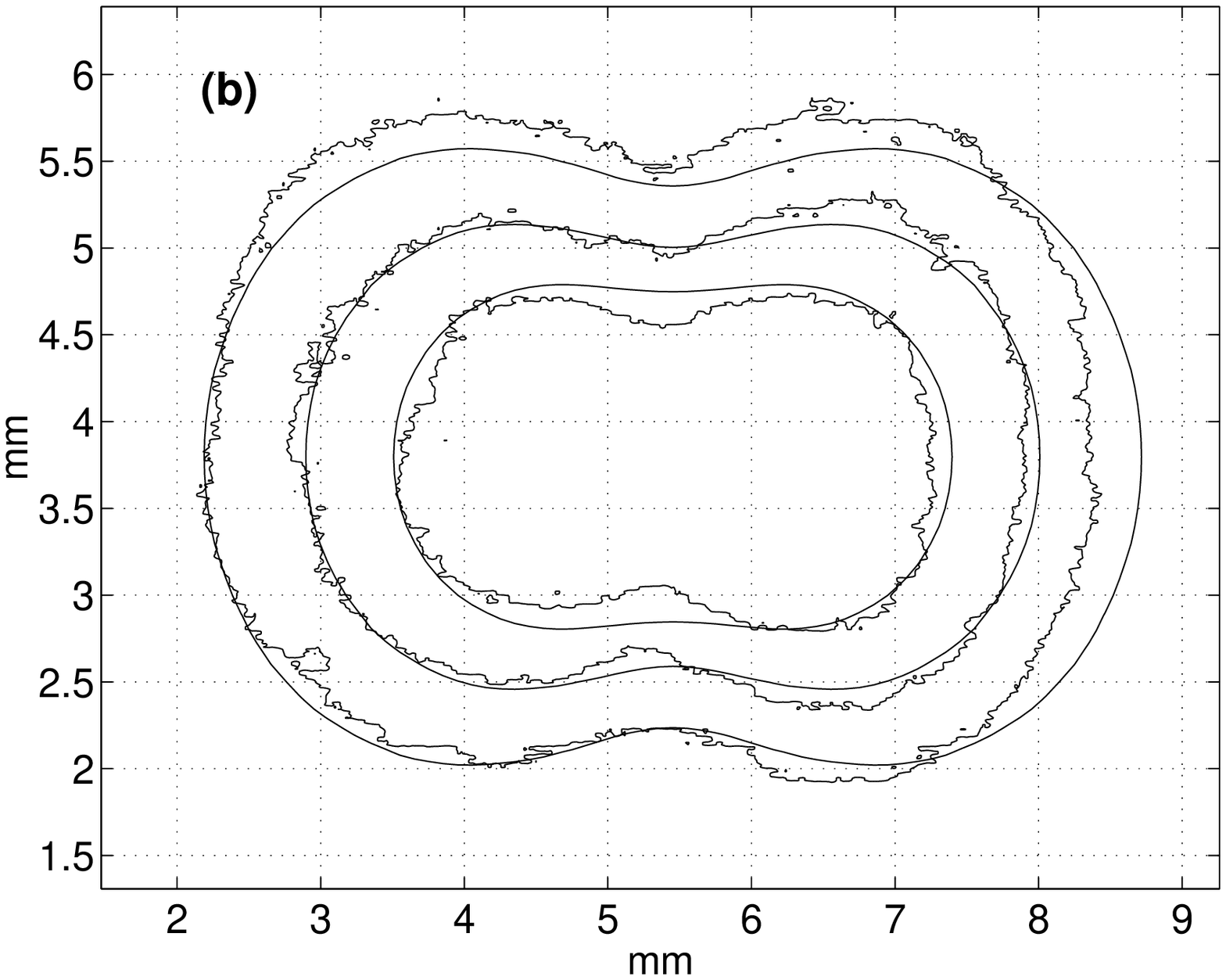,width=8cm,height=8cm}}
\centerline{\psfig{figure=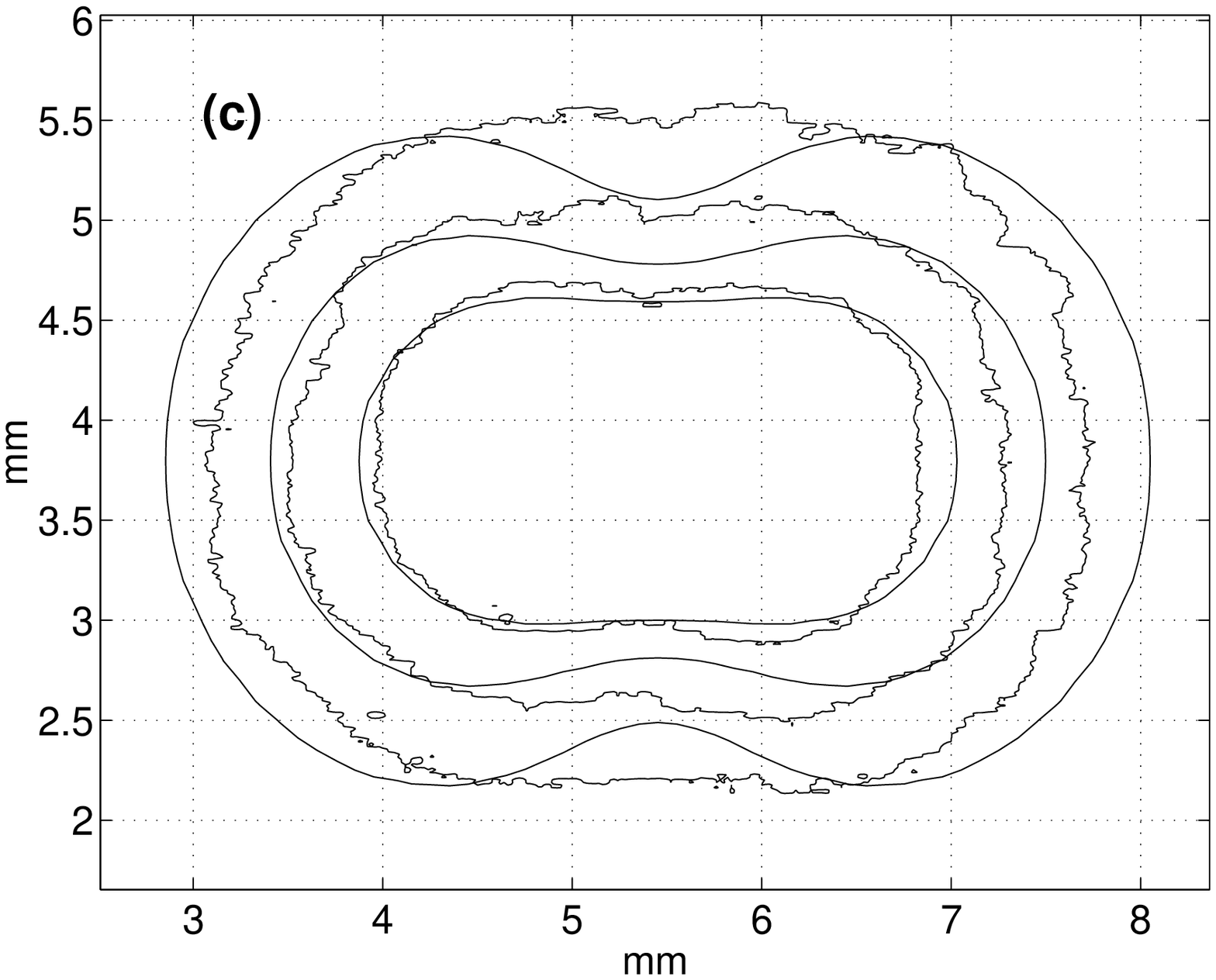,width=8cm,height=8cm}}
\caption{Comparison of experimentally obtained backscattering intensity
contours with the contours calculated by the slice model. The
scattering is in the PO geometry. The smooth lines are the calculated
contours while the jagged lines are experimental data. The figures
represent optical densities of $\tau$ = (a) 1.84, (b) 3.56, and (c)
8.76. respectively. Due to very strong stray light scattering, PO
contours for $\tau$ = 0.51 were not recorded.}
\end{figure}
At large $\tau$, the cusps show clear deviations from the calculated
intensity contours as can be seen in Figs. 9(d) and 10(c). We
believe that this again is because the depolarisation has not been
correctly accounted for. Further work is in progress to understand
the process of depolarisation in greater detail.

\section{Errors}

While the slice model provides excellent agreement with experiment in
the VH and PO geometries, it cannot calculate the intensity contours
correctly in the VV geometry. Once again, it is the depolarisation
that is responsible for this mismatch between theory and experiment.
While photons get depolarised as they travel into the medium, they are
obviously also depolarised as they scatter out. Figure 11(a) shows
superimposed calculated intensity contours at three intensity values,
for the PO and VV geometries. The dashed line is the total
backscattered intensity that is viewed in the PO geometry while the
solid lines are with the presence of an analyser oriented along the
$x$ axis. Experimentally, the shapes (but not the intensity values)
of the VV and PO contours are almost identical. For the outermost
contour, one can see that the only difference between the VV and PO
contours is their spread in the $x$ direction. In this region between
the two outermost contours, if the scattered photons are depolarised,
then the two contours will have nearly identical shapes. This
depolarisation does not however contribute significantly to the VH
contours because, as shown in Fig. 11(b), the region in which the PO
contours are intense is also the region where the depolarised source
scattering has its largest contribution. Thus, the effects of the
depolarisation of the outgoing scattered beam and that due to
scattering of the already depolarised incoming beam are hard to
separate as they scatter into almost the same region of space.

\begin{figure}[!h]
\centerline{\psfig{figure=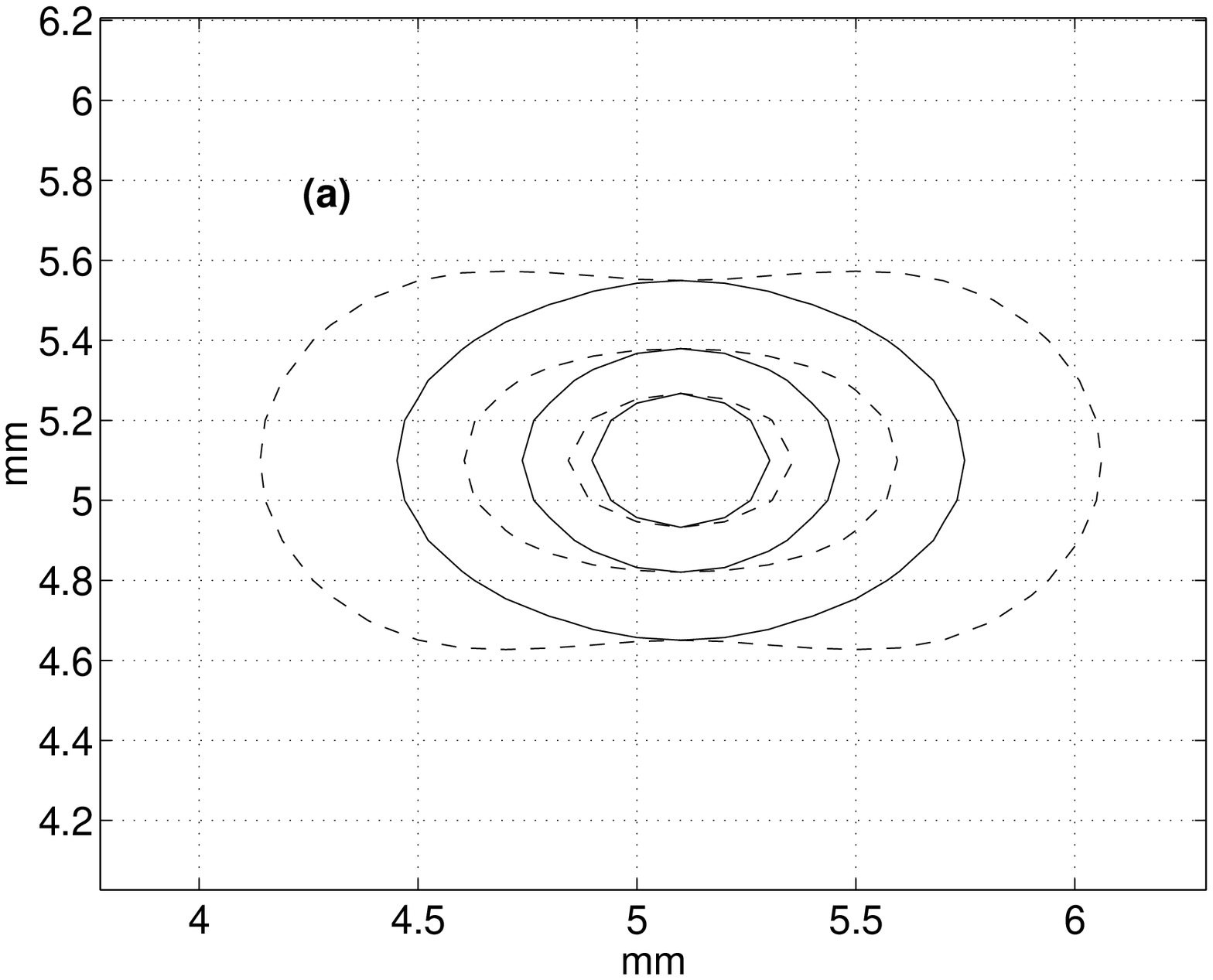,width=8cm,height=8cm}
\psfig{figure=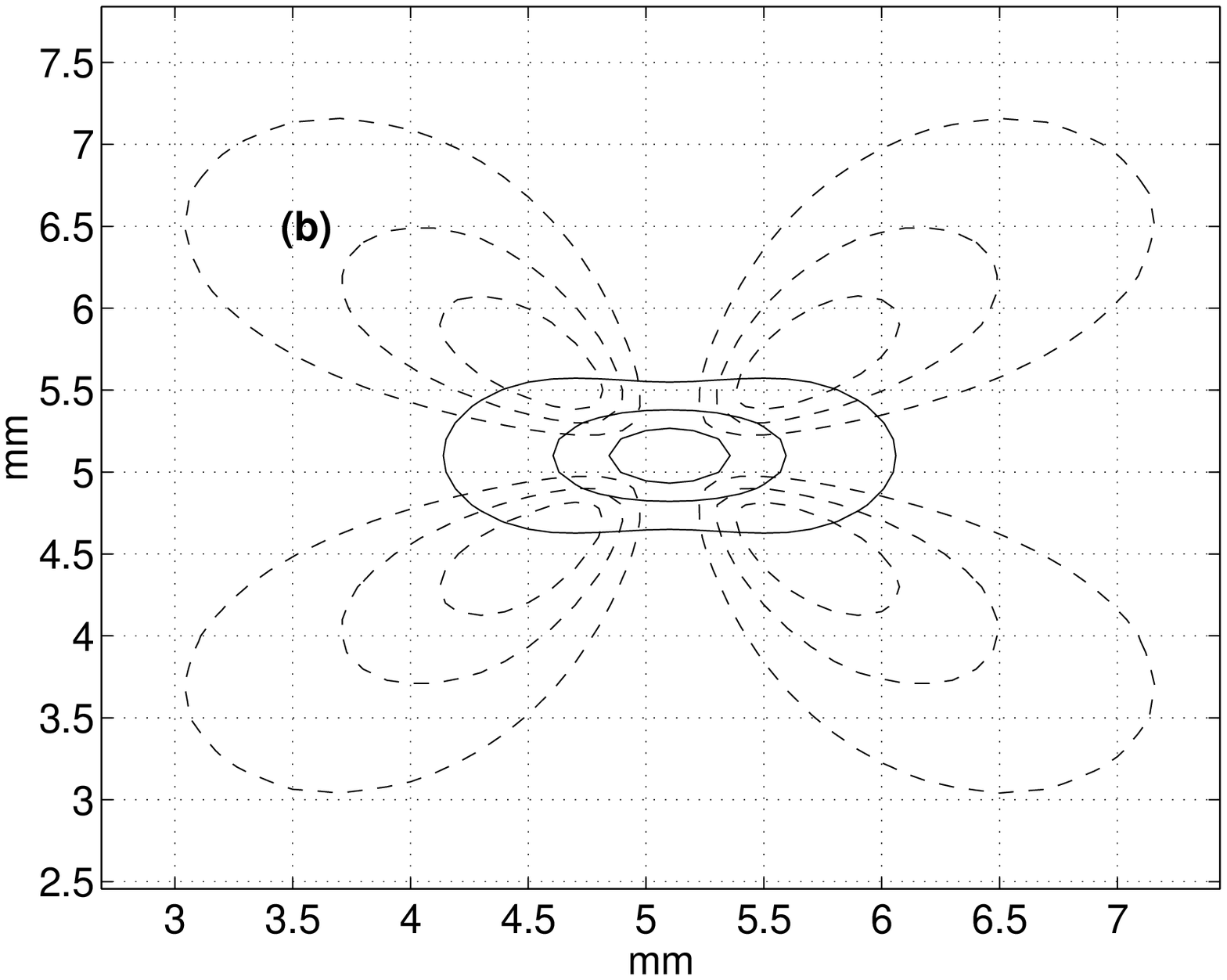,width=8cm,height=8cm}}
\caption{Calculated intensity contours at three intensity values are
superimposed and compared. Figure 11(a) compares PO and VV contours
while fig. 11(b) compares PO and VH contours. The depolarised and
diffuse fluxes are not included in these contours.}
\end{figure}
Interestingly, this indicates that when used as a tool to probe the
structure of the scatterers, it is best to use the VH patterns as a
probe, for three reasons. For one, the VH patterns are the least
vulnerable to depolarisation effects at the edges since the main
modifications of the patterns due to depolarisation are at the centre
as demonstrated in Fig. 11. For another, the VH patterns are usually
almost twice as large as the VV or PO patterns, thus providing a more
accurate determination of the contours, and finally stray reflections
of the direct beam are minimized when viewed with a crossed polariser.

Another shortcoming, as we have said earlier, is the absence of a
model to calculate depolarisation as a function of distance travelled.
Therefore, the model sometimes cannot reproduce the cusps in the
patterns correctly. Monte Carlo simulations which employ the full Mie
scattering cross sections are being carried out to improve our
understanding of the process of depolarisation.

\section{On the location of the effective source of diffusing photons}

When a beam of photons enters a scattering medium, photons are
scattered out of the beam and in due course, all the photons are
travelling random paths uncorrelated with one another. The transport
of the photon density within the medium can then be accurately
described by a diffusion equation. Often, the source of these
diffusing photons is modelled as a delta function. When the medium is
sufficiently optically `thick' ($L \sim 8l^{*}$), the assumption that
the delta function source lies at a depth of approximately one
transport mean free path inside the medium is a good approximation
\cite{Pine}. This depth is termed the penetration depth. It has
however been realised that if the region of validity of the diffusion
approximation is to be extended, then the penetration depth must be
more clearly understood \cite{Durian2}. Durian assumed an
exponentially distributed diffusing source but without achieving
considerable improvement. As we have seen, the production of
diffusing photons is described by the function ${\bf P}(r)$, a
function that is distinctly non-exponential function at small $r$ and
in fact, we have found that {\bf P(r)} is best fitted by a gamma
distribution. It is therefore unreasonable to expect the diffusion
approximation to provide a good approximation at small $r$.
\begin{figure}[!ht]
\centerline{\psfig{figure=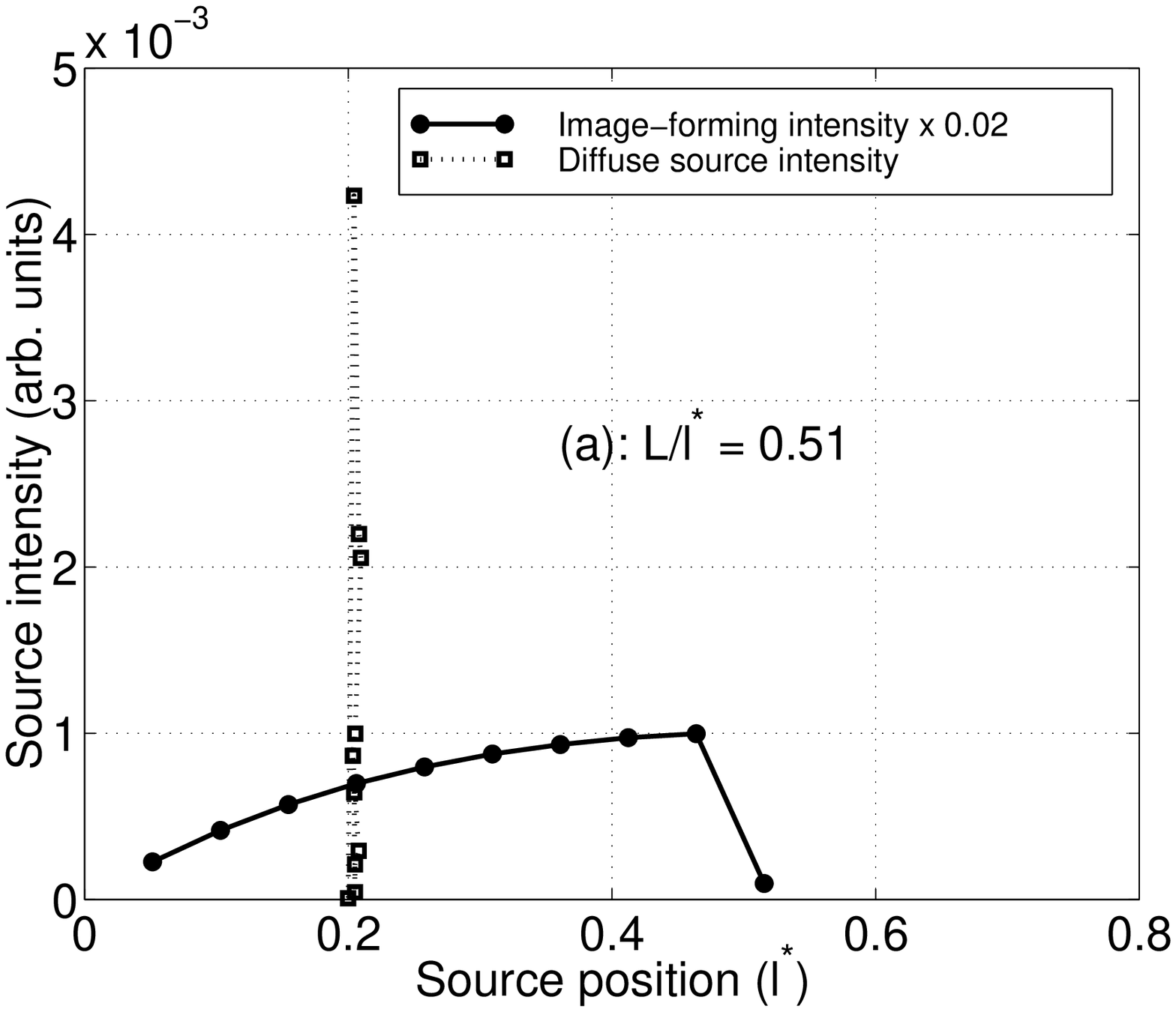,width=8cm,height=8cm}
\psfig{figure=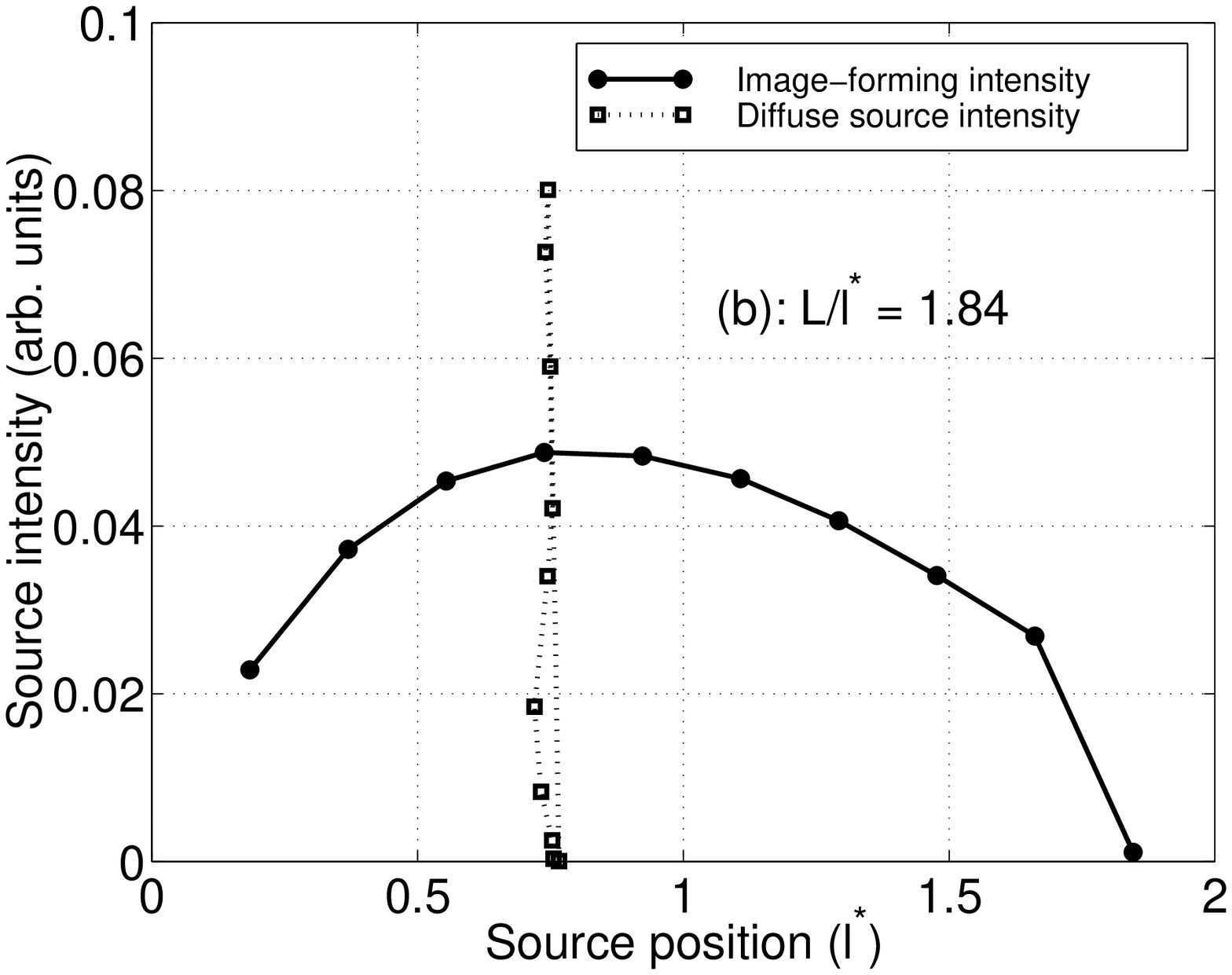,width=8cm,height=8cm}}
\centerline{\psfig{figure=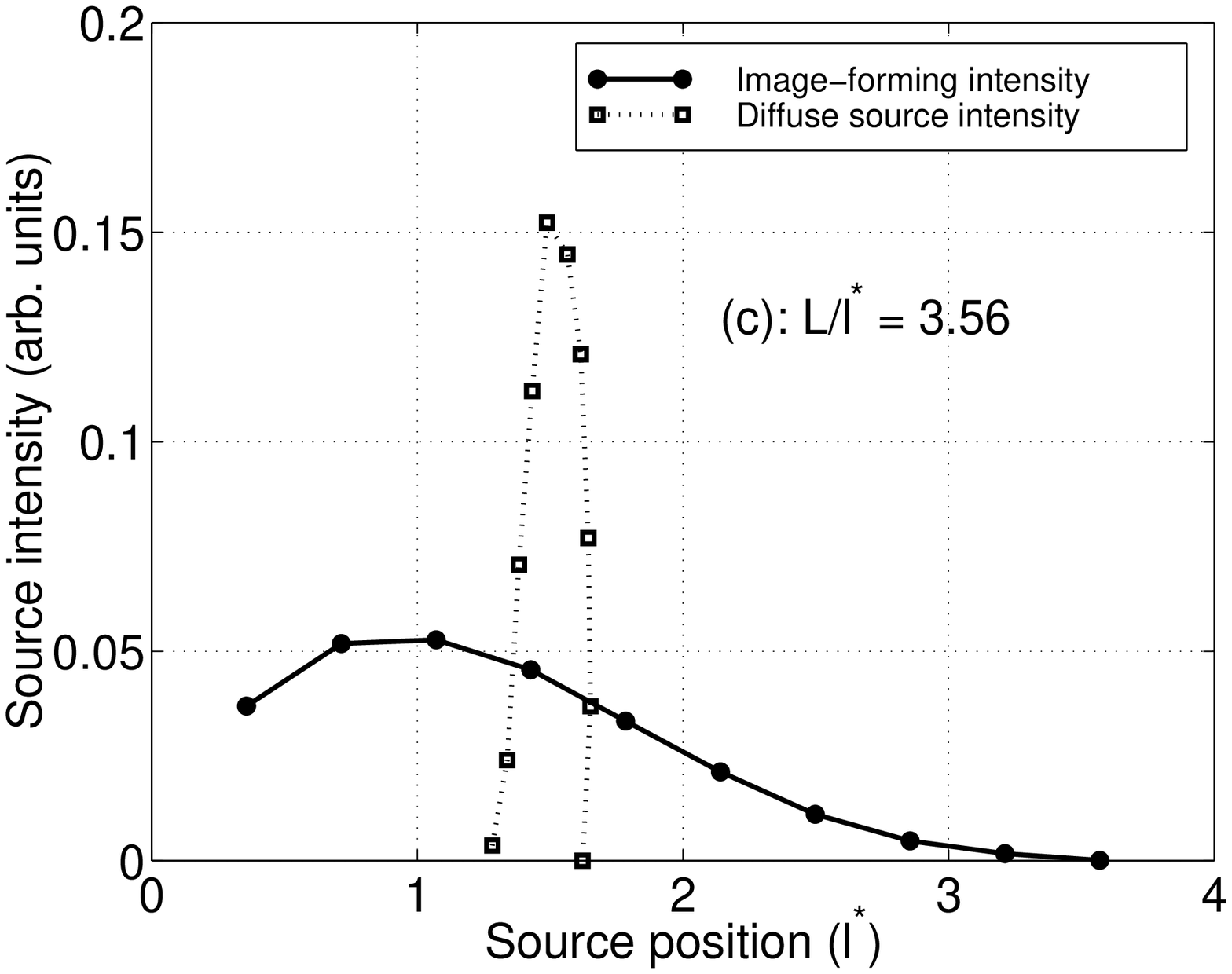,width=8cm,height=8cm}
\psfig{figure=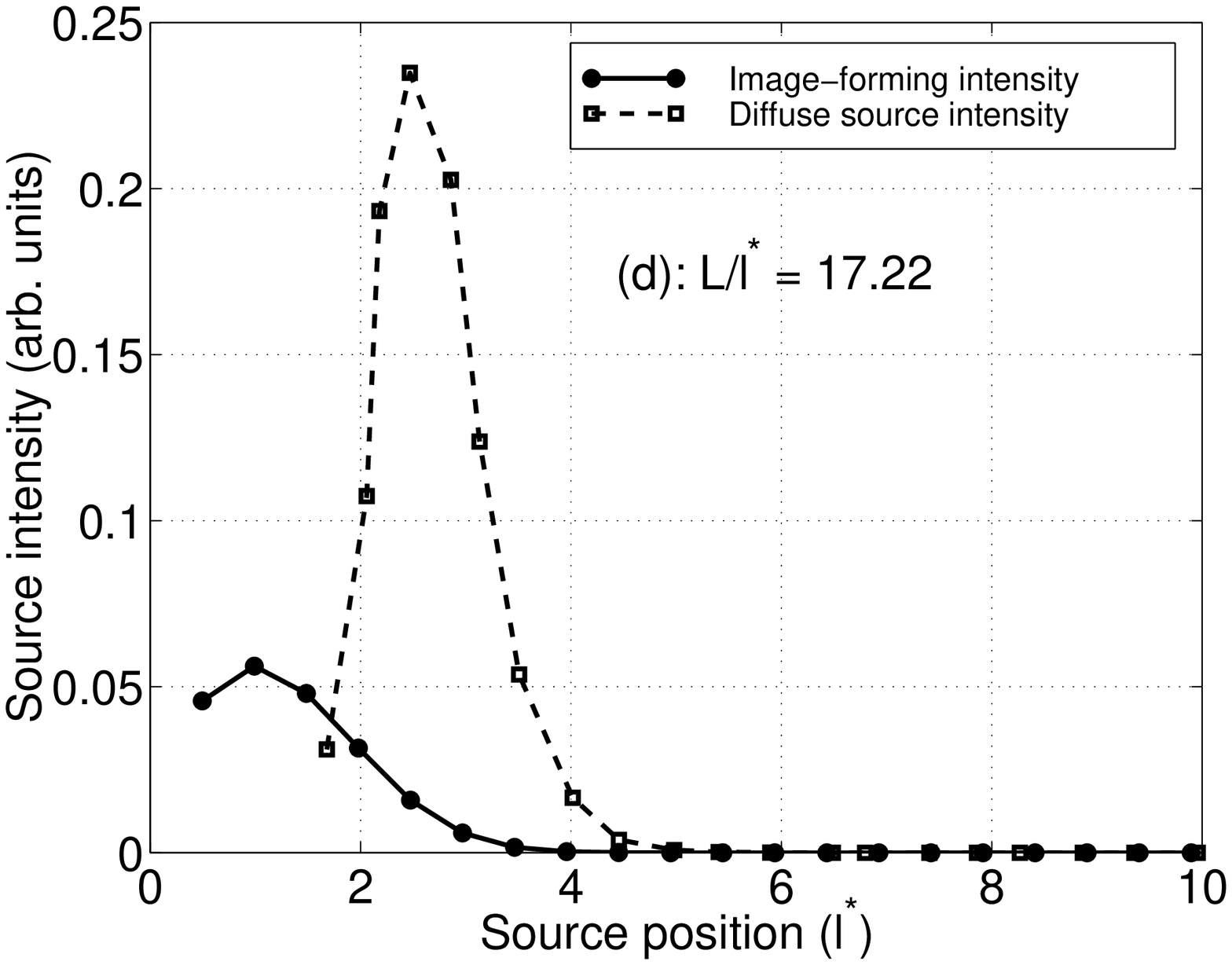,width=8cm,height=8cm}}
\caption{Intensities, measured in arbitrary units, of the diffuse and image
forming sources in slabs of varying optical densities as calculated by
the slice model are shown. The optical densities are $\tau = $
(a)0.51, (b) 1.84, (c)3.56 and (d)17.22. In (a), due to the large
difference between the intensities of the diffuse and image-forming
sources, the intensities of the image-forming sources have been scaled
down by a factor of 0.02.}
\end{figure}
In the slice model, we have explicitly calculated the image-forming
and diffuse intensity components at each slice. Figures 12(a)-(d)
trace the evolution of the image-forming and diffuse source
intensities as the optical thickness of the sample is increased. At
small values of $\tau$, the source is well modelled by a delta
function at approximately the centre of the slab. With increasing
optical thickness, the source spreads out until it finally appears as
in Fig.12(d), for $\tau = 17.22$. In a recent publication, Kostko and
Pavlov \cite{Kostko} found evidence to believe that the diffuse source
lay at a depth $\Delta = 4.6l^{*} \pm 0.63l^{*}$. Our results also
point in the same direction, in that the effective source of diffusing
photons lies much deeper in the medium than the commonly assumed
penetration depth of one transport mean free path.

Additionally, the figures provide us with quantitative information of
the utility of the patterns as a diagnostic tool. From Fig. 12(d) we
can see that the last of the pattern forming intensity in the medium
arises at a depth of $\sim 4l^{*}$. On the return path, many of these
photons would be randomised and so it appears that beyond this depth,
the patterns are not of much use unless one is able to use some method
such as polarisation discrimination \cite{Ramachandran,Venkatesh2} to extract
polarisation preserving reflected snake photons emerging from greater
depths within the medium.

\section{Possible applications and conclusion}

In the light of our results, we discuss here a few possible
applications for the azimuthal patterns where our model might be
successfully employed to make non-invasive measurements to determine
structural information about the scattering medium. We have seen that
in an experimental arrangement such as the one we have employed, the
patterns probe, at best, a depth of about $4l^{*}$ into the medium.
Therefore, in densely scattering media, we are limited to close
subsurface analysis. The patterns are most informative when the
medium is weakly scattering with an optical depth of about $4l^{*}$ or
less. One important application that we foresee is imaging within the
eye. These patterns are sensitive detectors of changes in tissue
structure and might well be capable of detecting sensitively, the
onset of defects linked to aggregations in the eye such as cataracts.

In conjunction with dynamic light scattering (DLS), the patterns may
be used to determine particle size distributions. We feel that the
use of these patterns is superior to merely using DLS, since the
scattering patterns are a direct probe of the shape of the scatterers.
For example, scattering from cylinders is very different from
scattering from spheres. They could thus form novel probes to monitor
aerosols and particulate flows. The first steps in this direction
have already been taken. Kadoma and van Egmond \cite{Kadoma} and Liu
and Pine \cite{Liu} have used these patterns to monitor structural
changes in micellar suspensions under shear in real time. Much
potential exists in the use of these patterns for industrial flow
monitoring applications. One can easily conceive of air or water
pollution monitors that provide continuous visual information on
suspended particulate matter in the flow. However, to broaden the
scope of our model, we need to recalculate many of the parameters like
${\bf P}(r)$ and the offset length for other scatterers like cylinders
and spheroids for which a more general T-matrix calculation will have
to be employed \cite{Barber}.

In summary, we have presented a quantitative model to analyse the
azimuthal variations in the backscattered intensity scattered from an
aqueous suspension of spherical scatterers. To our knowledge, this is
the first such quantitative model in the literature. Our analysis of
the effect, in terms of the polarisation preserving `reflected snake
photons' is also new and entirely different from those previously
published. We have experimentally recorded these patterns in
scattering from aqueous suspensions of colloidal particles over a wide
range of optical densities. Results of our calculations are in
excellent agreement with our experimental data. In addition, our
simulations have helped deepen our understanding of the long standing
problem of the penetration depth for diffusing photons in a turbid
medium. We have presented a new result, which again to the best of
our knowledge has not been reported previously, on the position and
shape of the apparent source of diffusing photons in a random medium.
Finally, we have shown that these patterns have a probing depth of
approximately $4l^{*}$ within a scattering medium and with this in
mind, we have described a few potential applications.

\begin{center}
{\Large Acknowledgements}
\end{center}
We thank the Supercomputer Education and Research Centre (SERC) at the
Indian Institute of Science for computational facilities. AKS thanks
the Raman Research Institute for a visiting professorship. VG thanks
Rajaram Nityananda for helpful discussions.
AKS and VG\footnote{Author for correspondence,
e-mail : vgopal@physics.iisc.ernet.in}
thank the Board of Research in Nuclear Sciences, India, for financial
assistance.


\begin{thebibliography}{10}

\bibitem{Bigio1}
A.H. Hielscher, J.R. Mourant and I.J. Bigio, Appl. Optics, {\bf 36}, 125 (1997).

\bibitem{Bigio2}
A.H. Hielscher, A.A. Eick, J.R. Mourant, D.Shen, J.P. Freyer and I.J. Bigio. 
Optics Express, {\bf 1}, 441 (1997).

\bibitem{Johnson}
T.M. Johnson and J.R. Mourant.
Optics Express, {\bf 4}, 193 (1999).

\bibitem{Dogariu}
M. Dogariu and T. Asakura, Optical Engg., {\bf 35}, 2234 (1996).

\bibitem{Carswell}
A.I. Carswell and S.R. Pal, Appl. Optics, {\bf 24}, 3464 (1985)

\bibitem{Hochheimer}
B.F. Hochheimer and H. A. Kues, Appl. Optics, {/bf 21}, 3811 (1982).

\bibitem{Marston1}
P.L. Marston, Jl. Opt. Soc. Am., {\bf 73}, 1816 (1983).


\bibitem{Rakovic}
M.J. Rakovi\'{c} and G.W. Kattawar, Appl. Optics, {\bf 37}, 3333 (1998)

\bibitem{Morse and Feshbach}
P.M.Morse and H.Feshbach, {\it Methods of Theoretical Physics}, vol. I,
(McGraw Hill, New York, 1953). See sec. 2.4.

\bibitem{Bohren and Huffman}
C.F. Bohren and D.R. Huffman, {\it Absorption and Scattering of Light
by small Particles}, Chapter 4 and Appendix A, 
(Wiley Interscience, New York, 1983).
The codes in Appendix A are also available by anonymous ftp from :
ftp://astro.princeton.edu/draine/scat/bhmie

\bibitem{Feynman}
{\it The Feynman lectures on Physics}, Section 32-5, 
(Addison Wesley, New York, 1965).

\bibitem{lstar}
A simple derivation of the transport mean free path in analogy with the
persistence length of a polymer chain may be found on p   in :
D. J. Pine, D. A. Weitz, G. Maret, P. W. Wolf, E. Herbolzheimer and
P. M. Chaikin, {\it Scattering and Localization of Classical Waves
in Random Media}, (P. Sheng ed., World Scientific Publishing, (1990b)).

\bibitem{Pine}
D.J.Pine, D.A.Weitz, J.X,Zhu and E.Herbolzheimer,
J. Phys. (Paris), {\bf 51}, 2101 (1990).

\bibitem{Venkatesh2}
Venkatesh Gopal, S. Mujumdar, H. Ramachandran and A.K. Sood, (unpublished)
(can be downloaded from http://xxx.lanl.gov/abs/cond-mat/9906188)

\bibitem{Ping Sheng}
Ping Sheng,
{\it Introduction to Wave Scattering, Localization, and Mesoscopic Phenomena},
page 174, (Academic press, 1995)

\bibitem{Barber}
P.W. Barber and S.C. Hill, {\it Light Scattering by Particles : Computational
Methods}, (World Scientific, Singapore, 1990). We use the programs S2 and S3.

\bibitem{Numrec}
W.H.Press, S.A.Teukolsky, W.T.Vetterling and B.P.Flannery, {\it Numerical
Recipes in FORTRAN
- The art of scientific computing},(Cambridge 1992).

\bibitem{Durian1}
D. J. Durian, Phys. Rev. E {\bf 51}, 3350 (1995).

\bibitem{Kaplan}
P.D. Kaplan, M. H. Kao, A. G. Yodh and D. J. Pine,
Appl. Opt. {\bf 32}, 3828 (1993).

\bibitem{Prasad}
B. R. Prasad, H. Ramachandran, A. K. Sood, C. K. Subramanian and N. Kumar,
Appl. Opt. {\bf 36}, 7718 (1997).

\bibitem{Durian2}
D.J. Durian, Appl. Opt. {\bf 34}, 7100 (1995).

\bibitem{Kostko}
A.F. Kostko and V.A. Pavlov, Appl. Opt. {\bf 36}, 7577 (1997).

\bibitem{Ramachandran}
H. Ramachandran and A. Narayanan, Optics Comm. {\bf 154}, 255
(1998).

\bibitem{Kadoma}
I.A. Kagoma and J. van Egmond, Phys. Rev. Lett., {\bf 76}, 4432 (1996).

\bibitem{Liu}
C. Liu and D.J. Pine, Phys. Rev. Lett., {\bf 77}, 2121 (1996).

\end{thebibliography}
\end{document}